\newcommand{\diracslash}[1]{#1\llap{/\kern2pt}}

\newcommand{\be}{\begin{equation}}
\newcommand{\ee}{\end{equation}}
\newcommand{\bea}{\begin{eqnarray}}\index{\footnote{}}
\newcommand{\eea}{\end{eqnarray}}
\newcommand{\ba}[1]{\begin{array}{#1}}
\newcommand{\ea}{\end{array}}

\documentclass[12pt]{iopart}
\usepackage{setspace}
\usepackage{graphicx}
\usepackage[top=.5in, bottom=.5in, left=.8in, right=.8in]{geometry}
\begin{document}
\title{The effects of trap-confinement and interatomic interactions on Josephson effects and macroscopic quantum self-trapping for a Bose-Einstein Condensate}
\author{Abhik Kumar Saha$^1$, Kingshuk Adhikary$^1$, Subhanka Mal$^1$, Krishna Rai Dastidar$^1$ and Bimalendu Deb$^1$ }
  \address{$^1$ School of Physical Sciences, Indian Association for the Cultivation of Science, Jadavpur, Kolkata 700032, India.}
\begin{abstract}
We theoretically study the effects of trap-confinement and interatomic interactions on Josephson oscillations (JO) and macroscopic quantum self-trapping (MQST) for a Bose-Einstein condensate (BEC) confined in a trap which has a symmetric double-well (DW) potential along z-axis and 2D harmonic potentials along x- and y-axis. We consider three types of model interaction potentials: contact, long-range dipolar and finite-range potentials. Our results show that by changing the aspect ratio between the axial and radial trap sizes, one can induce a transition from JO to MQST for contact interactions with a small scattering length. For long-range dipolar interatomic interactions, we analyze transition from Rabi to Josephson regime and Josephson to MQST regime by changing the aspect ratio of the trap for a particular dipolar orientation. For a finite-range interaction, we study the effects of relatively large scattering length and effective range on JO and MQST. We show that JO and MQST are possible even if scattering length is relatively large, particularly near a narrow Feshbach resonance due to the finite-range effects.     
\end{abstract}
\newpage

\section{Introduction}
Josephson effect (JE), predicted more than half-a-century ago by Brian D. Josephson \cite{Joseph-1}, represents an unambiguous manifestation of macroscopic quantum effects. The main feature of this effect is that the electrons in cooper pairs can execute perpetual tunneling without any dissipation between two superconductors when the barrier between the superconductors is thin enough (typically the thickness is less than 10 $\mathring{A}$) \cite{book:josephson}. This happens because the two macroscopic wave-functions of the superconductors on both sides of the barrier overlap in the classically forbidden region inside the barrier. Though the theory of the Josephson junction was originally developed in the context of superconductivity, it can be applied as well to the physical systems with weakly coupled macroscopic wave-functions. JE with ultra-cold atoms was first proposed by Javanainen in 1986 \cite{Joseph-2}. After the experimental realization of BEC in 1995 \cite{Joseph-3}, JE with atomic condensate has attracted renewed interests, giving rise to new effects such as MQST which was first predicted by Smerzi and collaborators \cite{Joseph-4}. MQST has no analogue in a superconducting Josephson junction. In 2001 JE was observed in an array of Bosonic Josephson junctions (BJJ) \cite{Joseph-5}. Both JO and MQST were experimentally demonstrated in a single BJJ \cite{albiez,gati:2006}. Over the years, several experimental and theoretical works \cite{theory:joseph,shenoy,theory:joseph1} have demonstrated many effects such as external JE \cite{external:joseph,external:joseph1}, internal JE \cite{internal:joseph,internal:joseph1}, coherent tunneling oscillations of interacting bosons \cite{jack:pra:1996,spagnolli}, collapse and revival of Josephson oscillations \cite{Joseph-7}, etc. In most of these works, JE has been studied by modeling the atom-atom interaction with the well-known zero-range contact potential which is valid for a small scattering length or for a weakly interacting system. Recently Spagnolli {\it et al.} \cite{spagnolli} have shown a transition from Rabi to plasma regime via tuning the $s$-wave scattering length from negative to positive value. This happens because the tunability of the scattering length through a magnetic Feshbach resonance \cite{rmp:2010} or any other means leads to the change in the effective atom-atom interaction which in turn affects the JE and related phenomena. We here assert that, even without altering the scattering length, it would be possible to obtain Josephson to plasma or MQST transition by changing the aspect ratio between the axial and radial trap frequencies. Because, it is well-known \cite{bec1d:2001,bec1d:1996} that the effective interaction can be drastically modified by appropriately changing the trap-confinement or aspect ratio, leading to confinement-induced resonances \cite{delta1d,confinement:1} even if the corresponding 3D free space interaction strength or the scattering length is small. In a dipolar BEC, by varying the geometry of the trap \cite{dipole-1}, dipole polarization axis \cite{dipolar} and shape of a dipolar BEC \cite{santos} one can change dipole-dipole interaction (DDI) from attractive to repulsive or vice versa. But by changing the aspect ratio of the trap, the transitions from Rabi to Josephson or Josephson to MQST regimes for a particular dipole orientation are not explored so far.

Here we investigate the effects of trap-confinement, the effective range and strength of atom-atom interactions on JO and MQST. We find that, by changing the aspect ratio of the trap one can bring about a transition from Josephson to MQST regime and the transition point depends on the strength of interaction. We consider a model finite-range interaction potential of Jost and Kohn \cite{Joseph-finite-11,Joseph-finite-12} for exploring JE and related phenomena in Bose-condensed atoms interacting with a finite-range and relatively large scattering length.  Near a Feshbach resonance, JO and MQST may be described considering Jost-Kohn interaction potential. Our results suggest that it may be possible to study Josephson effects and MQST near a narrow Feshbach resonance for which the effective range is very large or may even become negative \cite{ohara:prl:2012,pra:2013:hulet,range:2013,hutson:2014}. Since a Feshbach resonance occurs due to the existence of a quasi-bound or a quasi-molecular state, such studies will enable one to unravel hither-to unexplored effects of molecular regime on Josephson physics.

The paper is organized in the following way. In section \ref{1}, we analyze the method of constructing BJJ under two-mode approximation. In section \ref{2}, we present stationary and dynamical solutions of BJJ analytically. In section \ref{3} we present and discuss our results. In the end, we conclude in section \ref{4}.

\section{Bose-Einstein condensate in a double-well potential: Bosonic Josephson junction}\label{1}
The time evolution of the condensate wave function $\psi({\bf r},t)$ in a trap potential $V_{trap}{\bf (r)}$ at $T=0$ K satisfies the Gross-Pitaevskii equation (GPE)
\begin{eqnarray}
\hspace{-0.2in}
i\hbar\frac{\partial\psi({\bf r},t)}{\partial t} = -\frac{\hbar^2}{2m}\nabla^2\psi({\bf r},t)+V_{trap}{\bf (r)}\psi({\bf r},t)
+\int|\psi({\bf r^{'}},t)|^2\psi({\bf r},t)V_{int}{\bf (|r-r^{'}|})d{\bf r^{'}}
\end{eqnarray}
where $V_{int}{\bf (|r-r^{'}|)}$ represent the inter-atomic interaction between two particles. $m$ is the mass of an atom.
We consider a model trap potential \cite{jpb:kingshuk} of the form
\begin{eqnarray}
V_{trap}{\bf (r)}=V(\rho)+V(z)=\frac{1}{2}m{\omega_\rho}^2\rho^2+\frac{1}{2}\xi^2(z^2-\eta^2)^2\nonumber
\end{eqnarray}
which has harmonic oscillations along radial directions (x- and y-axes) and a symmetric DW along z-axis. Here $\rho^2=x^2+y^2$, $\omega_{\rho}$ is radial frequency, $z={\pm\eta}$ are the two minimum points where the 1D DW potential vanishes and the barrier height is $V_0=\frac{1}{2}\xi^2\eta^4$. So, the parameter $\xi^{2}$ has the dimension of energy-length$^{-4}$. If $V_{0}$ is much larger than the ground state energy of the DW potential then each well will almost behave like a harmonic oscillator having frequency $\omega_z=\frac{2\xi\eta}{\sqrt{m}}$. We write ${\bf |r|}=\sqrt{\rho^2+z^2}$. In the strong radial confinement regime we assume that all the atoms occupy the ground state of the radial harmonic potential. Then, integrating over the radial harmonic oscillator states, one can obtain an effective 1D Hamiltonian for the system. We solve for single-particle 1D eigen functions and eigenvalues numerically using the method of discrete variable representation (DVR). The lowest two energy eigen functions being quasi-degenerate in which atoms can occupy a ground \textquotedblleft band\textquotedblright in presence of particle-particle interactions. For symmetric DW, the lowest eigenstate $\phi_{s}(z)$  is space-symmetric $(\phi_{s}(z)=\phi_{s}(-z))$ and the other quasi-degenerate state $\phi_{as}(z)$ is anti-symmetric $(\phi_{as}(z)=-\phi_{as}(-z))$.

To reduce the 3D GPE in 1D form, we assume that in the radial direction the BEC is confined in ground state $\psi(\rho)=\exp[-\rho^2/2\lambda]/\sqrt{\pi \lambda}$ of the transverse trap and the condensed wave function $\psi({\bf r},t)=\psi_{1D}(z,t)\psi(\rho)$, where $\lambda=\frac{\omega_z}{\omega_{\rho}}$ is the aspect ratio of the trap. Here we have used $a_{z}=\sqrt\frac{\hbar}{m\omega_{z}}$ as the unit of length and $\hbar\omega_{z}$ as the unit of energy, and also in our subsequent discussions we will use these units if not otherwise specified. Under tight-binding approximation, one can form two mode basis states $\phi_{+}(z)=\frac{\phi_{s}(z)+\phi_{as}(z)}{\sqrt{2}}$, $\phi_{-}(z)=\frac{\phi_{s}(z)-\phi_{as}(z)}{\sqrt{2}}$. Let us consider $\phi_{+}(z)$ as the left-well localized state and $\phi_{-}(z)$ as the right-well localized state. To better understand the dynamical oscillations of two weakly linked BEC's, the time-dependent condensed wave-function $\psi_{1D}(z,t)$ can be written as a linear combination of two wave-functions which are localized in either site of the DW under two-mode approximation
\begin{eqnarray}
 \psi_{1D}(z,t)=\psi_1(t)\phi_{+}(z)+\psi_2(t)\phi_{-}(z)
\end{eqnarray}
The interesting relevant axial dynamics is determined by the wave-function $\psi_{1D}(z,t)$. Let us consider the time-dependent amplitudes are $\psi_1(t)=\sqrt{N_1(t)}\exp [i\theta_1(t)]$ and $\psi_2(t)=\sqrt{N_2(t)}\exp [i\theta_2(t)]$, $N_{1,(2)}$ is the number of atoms and $\theta_{1,(2)}$, the phase in the well left(right). The normalization of the total wave function $\psi_{1D}(z,t)$ is fixed by the total atom number $N$=$N_1$+$N_2$ and in order to fulfill the condition for a weak link, we have $\int\phi_{+}(z)\phi_{-}(z)dz<<1$.

Substituting the two-mode condensed wave-function $\psi_{1D}(z,t)$ in the time-dependent GPE and integrating over the spatial coordinates, we get    
\begin{eqnarray}
\hspace{-0.2in}
i\hbar\frac{\partial\psi_1(t)}{\partial t} &=& \left[E_{1}+U|\psi_1(t)|^2+U_i|\psi_2(t)|^2+K|\psi_2(t)|^2)\right]\psi_1(t)\nonumber\\ &+&\left[-J+2I|\psi_1(t)|^2+I|\psi_2(t)|^2\right]\psi_2(t)+K\psi_2(t)^2\psi_1^*(t)+I\psi_1(t)^2\psi_2^*(t)
\end{eqnarray}
similarly we get
\begin{eqnarray}
\hspace{-0.2in}
i\hbar\frac{\partial\psi_2(t)}{\partial t} &=& \left[E_{2}+U|\psi_2(t)|^2+U_i|\psi_1(t)|^2+K|\psi_1(t)|^2\right]\psi_2(t)\nonumber\\ &+&\left[-J+2I|\psi_2(t)|^2+I|\psi_1(t)|^2\right]\psi_1(t)+I\psi_2(t)^2\psi_1^*(t)+K\psi_1(t)^2\psi_2^*(t)
\end{eqnarray}
where,
\begin{eqnarray}
 J=-\int\left[\frac{\hbar^2}{2m}(\nabla\phi_{+}\nabla\phi_{-}) +\phi_{+}V_{dw}(z)\phi_{-}\right]dz\nonumber
 \end{eqnarray}
\begin{eqnarray}
 E_{1,2}=\int\left[\frac{\hbar^2}{2m}|\nabla\phi_{+,-}|^2+|\phi_{+,-}|^2V_{dw}(z)\right]dz\nonumber
\end{eqnarray}
with
\begin{eqnarray}
 U_{i j}=\int\int|\phi_{i}(z)|^2|\phi_{j}(z^{'})|^2V_{int}(|z-z^{'}|)dzdz^{'}\nonumber
\end{eqnarray}
where $i=\pm$ , $j=\pm$.
\begin{eqnarray}
 K=\int\int\phi_{\mp}^*(z)\phi_{\pm}^*(z^{'})\phi_{\mp}(z^{'})\phi_{\pm}(z)V_{int}(|z-z^{'}|)dzdz^{'}\nonumber
\end{eqnarray}
\begin{eqnarray}
 I=\int\int|\phi_{\pm}(z)|^2\phi_{\pm}^*(z^{'})\phi_{\mp}(z^{'})V_{int}(|z-z^{'}|)dzdz^{'}\nonumber
\end{eqnarray}
Here we have four possible interaction parameters, namely, the on-site interaction $U_{i j}=U_{+ +}=U_{- -} $ when $i=j$, the inter-site interaction $U_{i}=U_{+ -}=U_{- +}$ when $i\neq j$, partial exchange interaction $I$ and exchange interaction $K$. $U_{i}$, $K$, $I$ are vanishingly small for a contact interaction, but they are finite for a long-range dipolar and finite-range interaction potential. Note that, all these interaction parameters are obtained as the matrix element of $V_{int}$ between the product of two single-particle wave-functions of a two-particle non-interacting system in the trap. However, for strong interaction or resonant interactions, one may require to use the wave-functions of an interacting pair of particles to calculate the interaction matrix elements of the model. We will come back to this issue later in section \ref{5:jost}.

To characterize the Josephson dynamics, we define the population imbalance $z(t)=\frac{N_1(t)-N_2(t)}{N}$ and phase difference $\theta(t)=\theta_2(t)-\theta_1(t)$. The system of equations governing the dynamics of the population imbalance $z(t)$, and phase difference $\theta(t)$ reads as 
\begin{eqnarray}
\dot{z}(t)=-{\sqrt{1-z^2(t)}}\sin[\theta(t)]+\widetilde{M}(1-z^2(t))\sin[2\theta(t)]
\label{e1}
\end{eqnarray}
\begin{eqnarray}
\dot{\theta}(t)=Mz(t)+\frac{z(t)}{\sqrt{1-z^2(t)}}\cos[\theta(t)]+\widetilde{M}z(t)\left[1-\cos[2\theta(t)]\right]
\label{e2}
\end{eqnarray}
where, we have rescaled to a dimensionless time $t(2J-2NI)/\hbar\rightarrow t$ and $M=\frac{(NU-NU_{i}-2NK)}{(2J-2NI)}$, $\widetilde{M}=\frac{NK}{(2J-2NI)}$. The dimensionless parameters $M$ and $\widetilde{M}$ determines different dynamical regimes of the BEC atomic tunneling. If we neglect all interaction terms except the on-site interaction, then the Eq.(\ref{e1}) and (\ref{e2}) reduce to the form of BJJ with contact interaction \cite{shenoy}. In that case, if we change the sign of $U$ from repulsive to attractive, then to maintain symmetry between Eq.(\ref{e1}) and (\ref{e2}) we have to change the phase $\theta\rightarrow\pi-\theta$. So the dynamics of BJJ remains same in attractive interactions with a phase shift of $\pi$. But in the case of long-range and finite-range interactions, if we change the dimensionless interaction parameters from positive to negative value we can not recover the symmetry again with any change of phase shift. This is because of the interaction term $\widetilde{M}$. Thus the dynamics of the system with a long- or finite-range interaction is completely different when we go to the repulsive to attractive interaction.
The two-mode Hamiltonian can be written in the form
\begin{eqnarray}
H =(M+\widetilde{M})\frac{z^2}{2}-\sqrt{1-z^2}\cos\theta+\frac{\widetilde{M}}{2}(1-z^2)\cos2\theta
\end{eqnarray}
In the following we will restrict the discussion of the Josephson dynamics to the case of a symmetric DW potential with $E_{1}=E_{2}$ and equations of motion can be written in Hamiltonian form $\dot{z}=-\frac{\partial H}{\partial\theta}$, $\dot{\theta}=\frac{\partial H}{\partial z}$. Therefore $z$ and $\theta$ are the canonically conjugate variables.
\section{Solutions}\label{2}
\subsection{Stationary solutions}
We get stationary solutions by setting $\dot{z}$ and $\dot{\theta}$ equal to zero \cite{shenoy}. These are $z_{s}=0$, $\theta_{s}=2n\pi$ in which eigen energy is $E_{-}=-1+\frac{\widetilde{M}}{2}$. The next stationary state is $z_{s}=0$, $\theta_{s}=(2n+1)\pi$ with eigen energy $E_{+}=1+\frac{\widetilde{M}}{2}$. In the case of non-interacting atoms in a symmetric DW potential, the eigenstates are always symmetric or antisymmetric with $z_{s}=0$. But due to the nonlinear interaction there is a symmetry breaking in $z$ corresponding to  $\theta_{s}=(2n+1)\pi$ and $z_{s}=\pm\sqrt{1-\frac{1}{M^2}}$, provided $|M|>1$ with energy $E=\frac{1}{2}\left[M+\widetilde{M}+\frac{1}{M}\right]$.

\subsection{Stability analysis of stationary solutions}
The linear stability of the condensate can be understood by analysing the equations $\frac{\partial H}{\partial z}{\Big|}_{z_{s},\theta_{s}}=0$; $\frac{\partial H}{\partial \theta}{\Big|}_{z_{s},\theta_{s}}=0$, where $H$ is the two mode Hamiltonian. Towards this end, we study the Hessian matrix of the system for $\theta_{s}=0$ and $\pi$. The Hessian matrix for this system is always diagonal and its eigenvalues are $\frac{\partial^2 H}{\partial z^2}{\Big|}_{z_{s},\theta_{s}}$ and $\frac{\partial^2 H}{\partial \theta^2}{\Big|}_{z_{s},\theta_{s}}$. The diagonal terms of the matrix are 
\begin{eqnarray}
\frac{\partial^2 H}{\partial z^{2}}=(M+\widetilde{M})-\widetilde{M}\cos2\theta+\frac{\cos\theta}{(1-z^2)^\frac{3}{2}}\nonumber
\end{eqnarray}
\begin{eqnarray}
\frac{\partial^2 H}{\partial \theta^2}=\sqrt{1-z^2}\cos\theta-2\widetilde{M}(1-z^2)\cos2\theta\nonumber
\end{eqnarray}
Depending on the sign of eigenvalues, a stationary point will be maxima, or minima or saddle point. In our case the eigenvalues depend on all interaction terms and the oscillations around a stationary point occur only if the stationary point is either minimum or maximum.

\subsection{Dynamical solutions}
For non-interacting atoms, we get sinusoidal oscillations which refer to as a Rabi oscillations in which population imbalance $z$ vary with time with frequency $\omega_{R}=\frac{2J}{\hbar}$. The two sets of stationary population imbalance and the phase are (i) $z=0$, $\theta=0$ and (ii) $z=0$, $\theta=\pi$. Now if we linearize the Eq.(\ref{e1}) and (\ref{e2}) around these values we get zero-phase Josephson frequency of the form 
\begin{eqnarray}
\omega_{0}=\sqrt{(1+M)(1-2\widetilde{M})}\nonumber
\end{eqnarray}
and $\pi$-phase Josephson frequency of the form
\begin{eqnarray}
\omega_{\pi}=\sqrt{(1-M)(1+2\widetilde{M})}\nonumber
\end{eqnarray}
The small amplitude oscillations frequency around zero-phase mode is applicable only when $M>-1$ and $\widetilde{M}<0.5 $. For $\pi$-phase mode it is applicable only $M<1$ and $\widetilde{M}>-0.5$ . The oscillation frequency in $\pi$-mode is always less than that of the zero-mode for positive $M$. But, for negative $M$, the oscillation frequency for $\pi$-mode is always greater than that of the zero-mode.

\subsection{Macroscopic quantum self-trapping}
The dynamics of the system changes drastically, when the initial population imbalance exceeds a critical value $z_{c}$. In that case, the tunneling is strongly suppressed, resulting in self-trapping. Once the critical value is reached, the tunneling current between two wells is such that the current gets reversed before even going to zero. This situation occurs because the phase exceeds $\pi$ before $\dot z$ goes to zero. Consequently, the population imbalance remains non-zero throughout a complete cycle. This is the case where the population is trapped in a single-well although there is a Josephson current between two wells. To evaluate the condition for MQST, the initial energy for $z(0)=z_{c}$ and $\theta(0)=0$ has to be large enough to reach $\theta_{final}=\pi $  at $z_{final}=0$ which correspond to an energy 
\begin{eqnarray}
 H_{final}=1+\frac{\widetilde{M}}{2}\nonumber
\end{eqnarray}

\begin{eqnarray}
H_{0}\equiv H(z(0)=z_{c},\theta(0)&=&0)=(M+\widetilde{M})\frac{z_{c}^2}{2}-\sqrt{1-z_{c}^2}\cos\theta(0)\nonumber\\&+&\frac{\widetilde{M}}{2}(1-z_{c}^2)\cos2\theta(0)\equiv 1+\frac{\widetilde{M}}{2}\nonumber
\end{eqnarray}
Using the above condition we find that the critical population imbalance for MQST for zero initial phase difference is given by
\begin{eqnarray}
z_{c}= \frac{2}{M}\sqrt{M-1}\nonumber
\end{eqnarray}
For zero initial phase difference, we get self-trapping only when $|M|>2$ and for small values of $M$ ($>2$) self-trapping occurs even at arbitrary large initial population imbalance $z(0)$. In order to reach self-trapping, one has to increase $z(0)$ above a critical value $z_{c}$ for fixed $M$ and $\widetilde{M}$, or alternatively increase $M$ by changing the interaction parameters or total number of atoms keeping $z(0)$ fixed. We get an expression for scaled critical interaction energy
\begin{eqnarray}
 M_{c}=\frac{2\left[1+\sqrt{1-z(0)^2}\cos(\theta(0))+\frac{\widetilde M}{2}[1-\cos(2\theta(0))]+\frac{\widetilde M z(0)^2}{2}[1-\cos(2\theta(0))\right]}{z(0)^2}
\end{eqnarray}
In the zero or $\pi$-phase mode it gives same result \cite{shenoy}. The MQST is a nonlinear effect arising from inter-particle interactions in the individual wells. It is self-maintained in a closed system without an external drive.

\section{Results and discussion}\label{3}
\subsection{Contact interaction}\label{5:contact}
We consider three types of interatomic interactions: contact, long-range dipolar and finite-range interaction potential. We first discuss the case of delta potential $V_{\delta}({\bf r})=g\delta({\bf r})$, where $g=\frac{2\pi\hbar^2a_{s}}{\mu}$, $a_{s}$ be the 3D $s$-wave scattering length and $\mu=\frac{m}{2}$ is the reduced mass. For numerical illustration with realistic parameters, we consider a BEC of $^{39}$K, with $s$-wave scattering length $a_{s}=0.05$ nm. We fix the DW axial frequency $\omega_{z}=2\pi\times85$ Hz \cite{spagnolli}, $a_{z}=1740$ nm and set the minima of the DW potential at $\eta=\pm 2 a_{z}$, the barrier height is $V_{0}=2\hbar\omega_{z}$. By changing the radial frequency of the trap from large value to small value upto $\lambda\rightarrow1$, we wish to investigate how the interaction parameters change. It is worth-mentioning that, by changing $\lambda$ from small to large values, one can effectively go from quasi-1D to isotropic 3D regime. We retain only three types of interaction parameters because the inter-site interaction and exchange interaction are same for a contact interaction. The value of the tunneling co-efficient is found to be $J=12.81$ Hz and total number of atoms is taken to be $N=1000$.

\begin{figure}[!htbp]
\centering
\vspace{.25in}
\begin{tabular}{@{}cccc@{}}
\hspace{-0.2in}
\includegraphics[height=2.2in, width=3in]{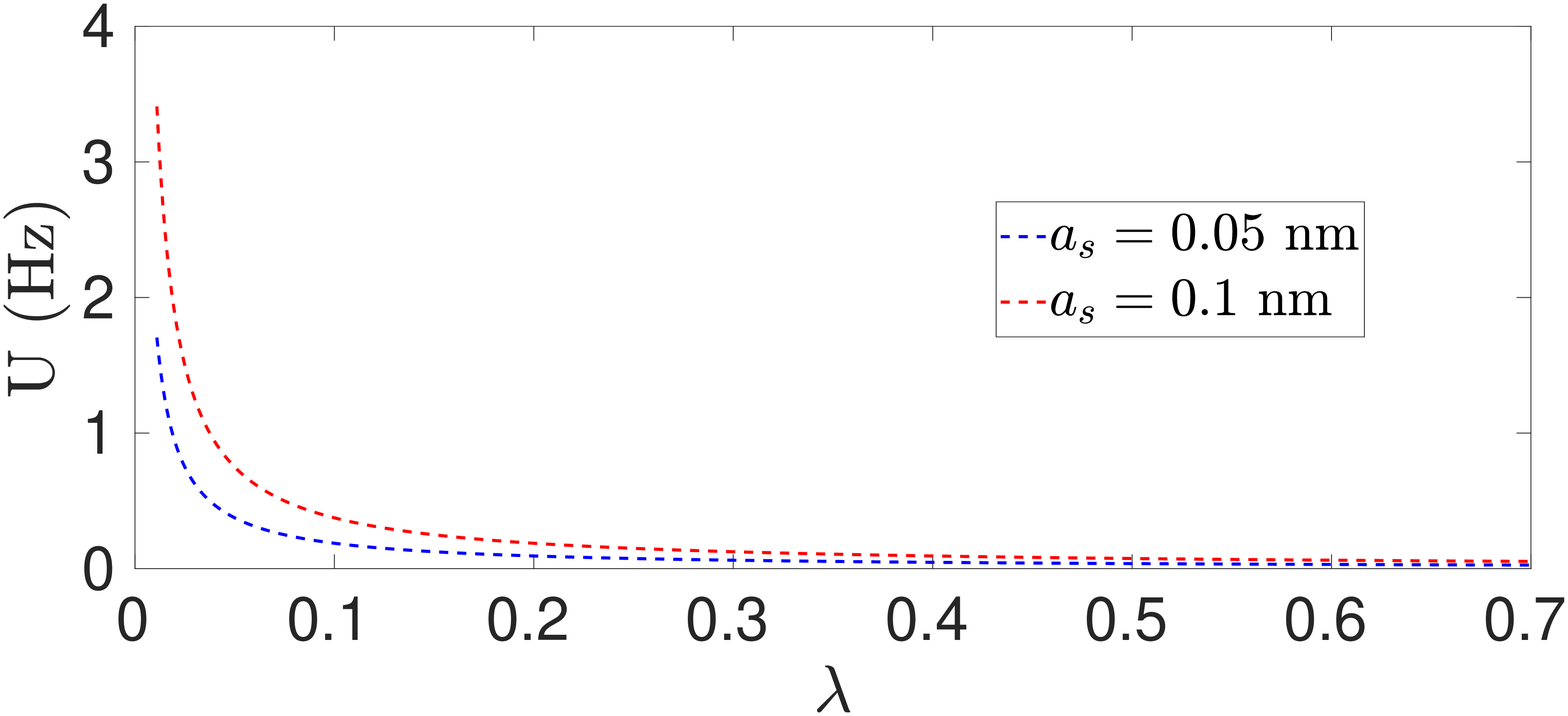} &
\includegraphics[height=2.2in, width=3in]{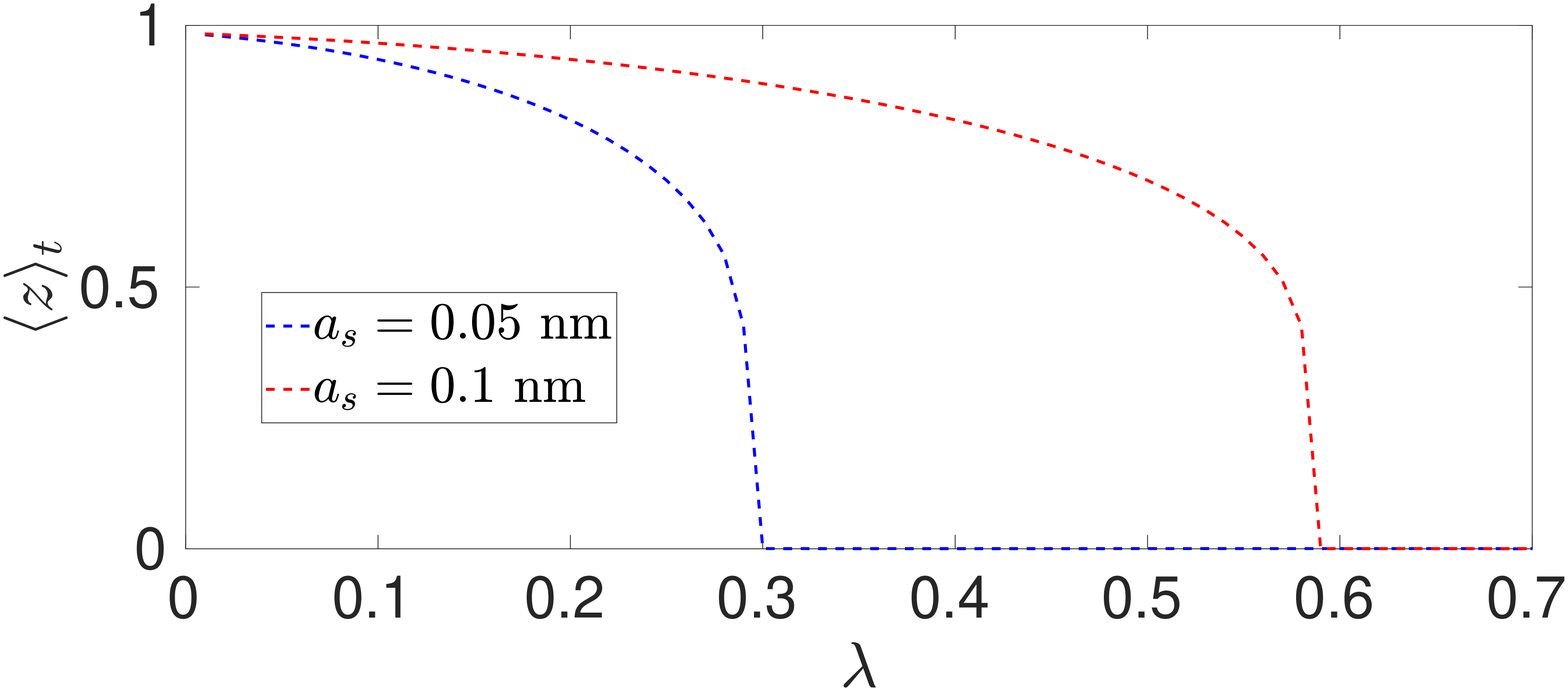}\\
\end{tabular}
\caption{\small Variation of on-site interaction energy $U$ (left) in Hz and time-averaged population imbalance $z(t)$ (right) as a function of aspect ratio $\lambda$ with total number of atoms $N=1000$.}
\label{Figure1}
\end{figure}

From the left side of Fig.\ref{Figure1}, it is clear that by changing the aspect ratio one can change the interaction energy keeping the scattering length $a_{s}$ and $N$ constant. We see that the value of the on-site interaction  is large in the quasi-1D limit. When the trap is almost isotropic ($\lambda \simeq 1$) we get a situation where $NU\approx J$ and in the quasi-1D limit ($\lambda <<1 $) $NU$ is much greater than $J$ leading to plasma regime \cite{Joseph-4}. We define time-averaged population imbalance $\langle z\rangle_{t}$ as the average of $z(t)$ over the time-period of oscillations. In the right side of Fig.\ref{Figure1} we plot $\langle z\rangle_{t}$ as a function of $\lambda$ for zero-phase mode. This figure shows that as $\lambda$ decreases below a critical value, the system undergoes a transition from JO ($\langle z\rangle_{t}=0$) to MQST ($\langle z\rangle_{t}\neq0$). From this figure we notice that the transition point moves towards larger value of $\lambda$ as $a_{s}$ increases.

To verify whether our quasi-1D contact interaction can give exact result to the 1D form of the regularized delta potential, we consider 1D form of the regularized delta potential \cite{delta1d} $V^{1D}_{\delta}(z)=g_{1D}\delta(z)$, where $g_{1D}=\frac{2\hbar^2a_{s}}{\mu a_{\rho}^2}\left(1-\frac{Ca_{s}}{a_{\rho}}\right)^{-1}$, $C$ is a constant, the value of $C=1.4603$ and $a_{\rho}=\sqrt\frac{\hbar}{m\omega_{\rho}}$ is the size of the transverse harmonic potential. We find that our effective quasi-1D interaction gives almost similar results as for 1D regularized delta potential interaction. 

\subsection{Dipole-dipole interaction}\label{5:dipole}
\begin{figure}[!htbp]
\centering
\vspace{.25in}
\begin{tabular}{@{}cccc@{}}
\hspace{-0.2in}
\includegraphics[height=2.2in, width=3in]{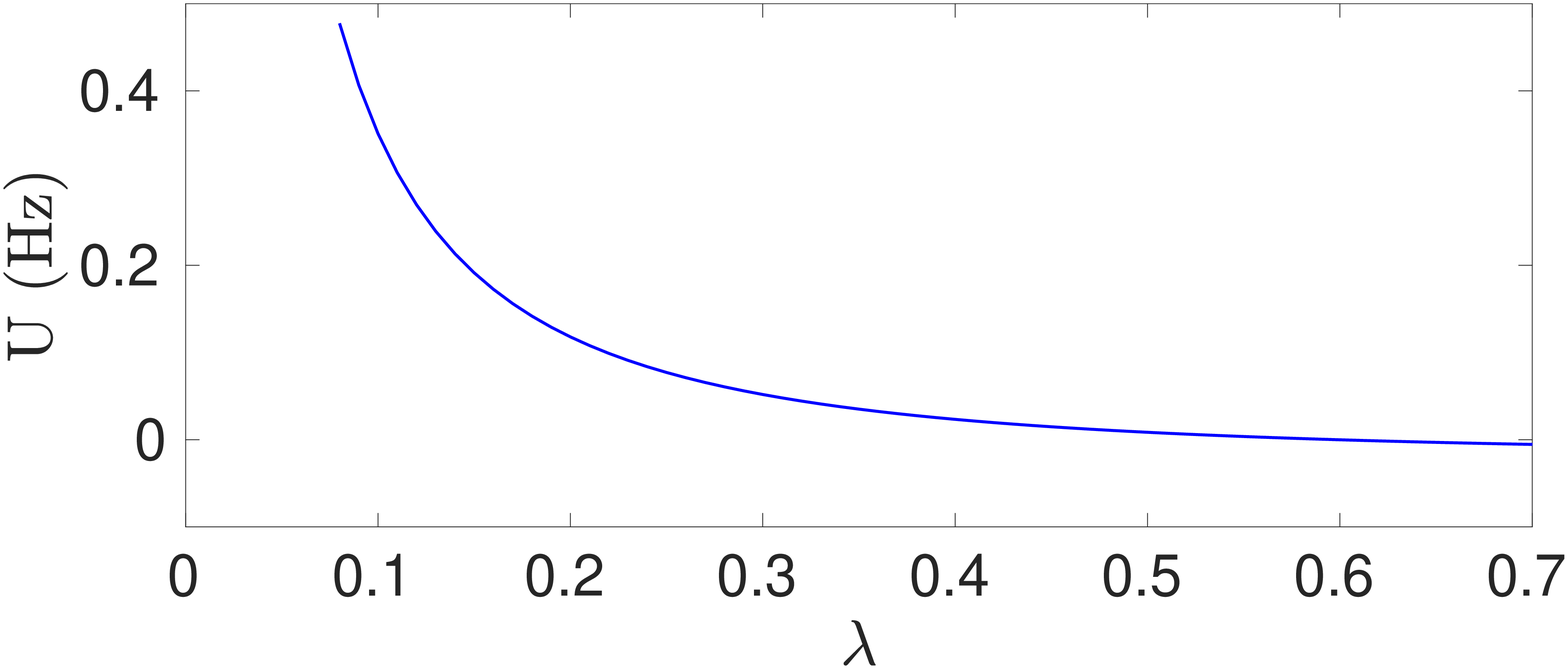} &
\includegraphics[height=2.2in, width=3in]{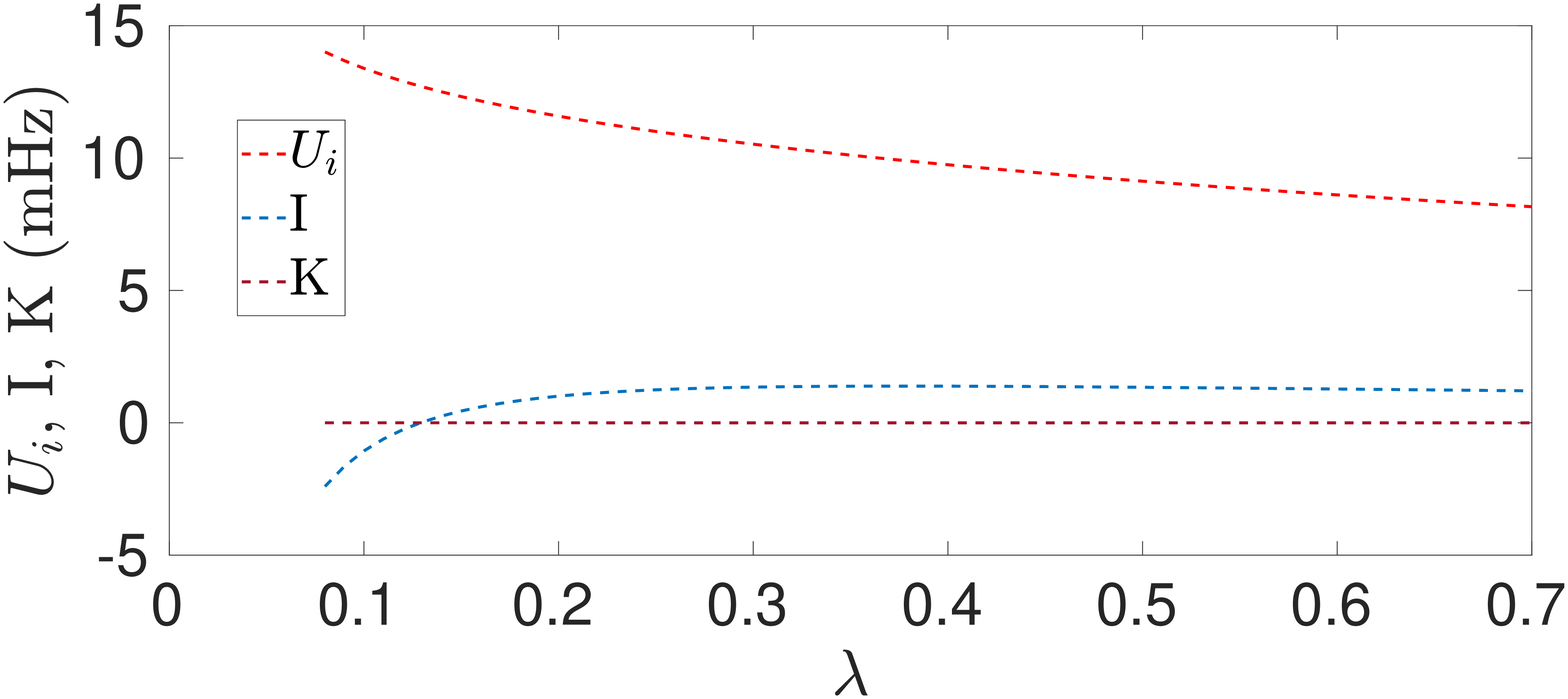}\\
\end{tabular}
\caption{\small Variation of $U$ (left) in Hz, inter-site interaction $U_{i}$, partial exchange interaction $I$, exchange interaction $K$ (right) in mHz  as a function of $\lambda$ for DDI.}
\label{Figure3}
\end{figure}

\begin{figure}[!htbp]
\centering
\vspace{.25in}
\begin{tabular}{@{}cccc@{}}
\hspace{-0.2in}
\includegraphics[height=2.2in, width=3in]{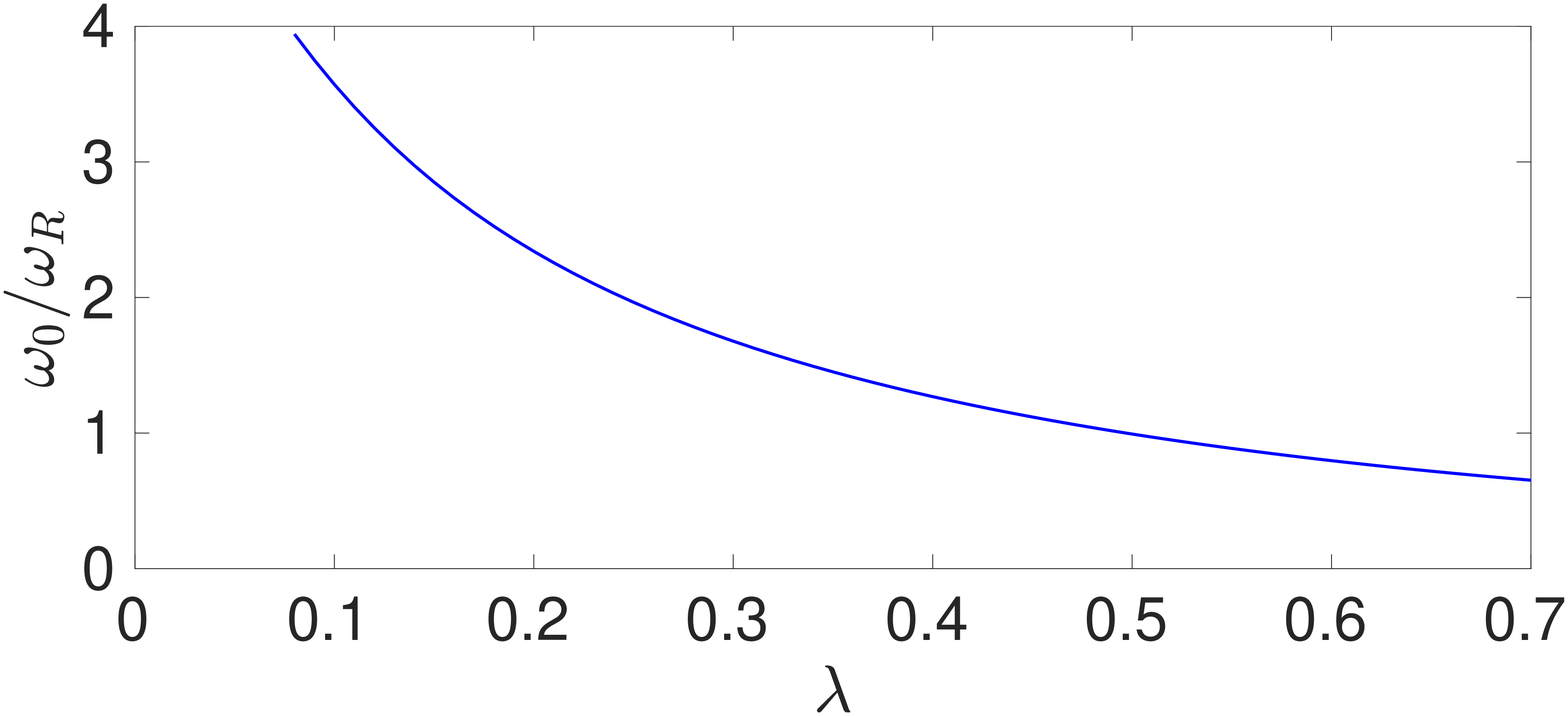} &
\includegraphics[height=2.2in, width=3in]{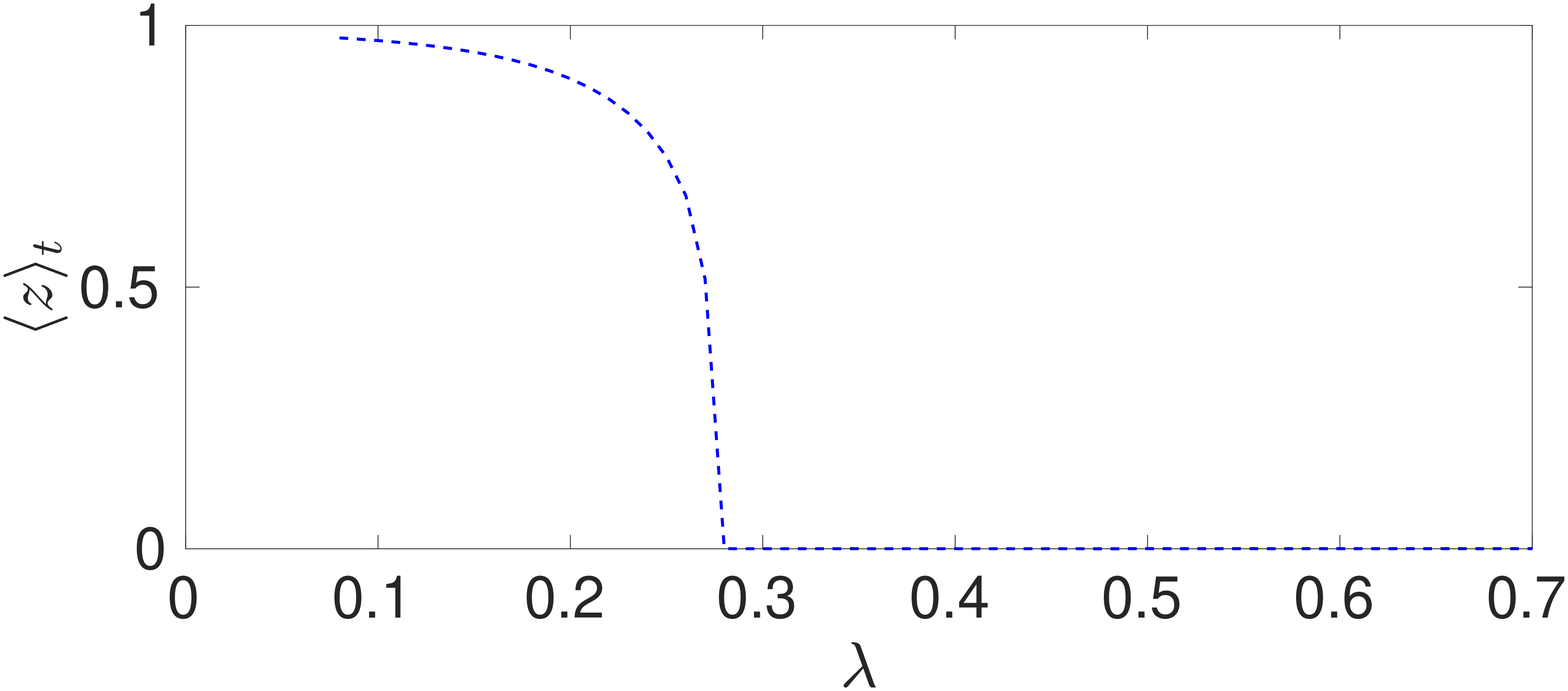}\\
\end{tabular}
\caption{\small Variation of Josephson frequency $\omega_{0}$ (in unit of $\omega_{R}$) (left) in the limit of small-amplitude oscillations and time-averaged population imbalance $z(t)$ (right) as a function of $\lambda$ with total number of atoms $N=1000$ for zero-phase mode for DDI.}

\label{Figure5}
\end{figure}

Next, we consider long-range dipolar potential. Here we consider $N$ number of dipolar bosons aligned in the y-z plane by external field. Then the DDI is given by
 \begin{eqnarray}
V^{dd}_{int}(|{\bf r}|) = \frac{\mu_{d}^2}{r^3}(1-3\cos^2\phi_{d})
 \end{eqnarray}
where $\mu_{d}$ is the magnetic dipole moment of each atom, $\bf r$ is inter-atomic distance, $\phi_{d}$ is the angle between ${\bf r}$ and polarization direction. The effective 1D form of the DDI \cite{adhikari} is given by
\begin{eqnarray}
 V^{1D}_{dd} &=& \frac{\mu_0\mu_{d}^2}{4\pi}\frac{(1+3\cos2\phi)}{8a_\rho^3}\Bigg\{\frac{8}{3}\delta\left(\frac{|z|}{a_\rho}\right)+\frac{2|z|}{a_\rho}\nonumber\\ &-&\sqrt{2\pi}\left(1+\frac{|z|^2}{a_\rho^2}\right)e^{|z|^2/2a_\rho^2} erfc\left(\frac{|z|}{\sqrt{2}a_\rho}\right)\Bigg\}
\end{eqnarray}
$|z|=|z-z^{'}|$, 1D inter-particle separation, $\mu_{0}$ is the permeability of free space, $erfc$ is the complementary error function and $\phi$ is the angle between polarized dipole orientation with $z$ axis. It is known that by changing the dipole orientation one can change the interaction. The transition from JO to the MQST has been studied by changing the dipole orientation with the polarized axis \cite{dipolar}. But here we find that one can also change or switch the interaction from repulsive to attractive by changing the aspect ratio of the trap for a fixed dipole orientation in order to obtain transition between JO and MQST. We consider a dipolar BEC of $^{52}$Cr, which has large magnetic moment $\mu_{d}\approx6\mu_{B}$ ($\mu_{B}$ is bohr magneton) and we assume that the short range forces do not affect the long-range DDI. For numerical illustration we keep the total number of atoms $N$ and DW trapping frequency same as used for contact potential interaction. Here we have to necessarily consider all four interaction terms. We fix the value of  $\phi=0.69\pi$ so that our effective on-site interaction $NU$ lies well below band gap of the DW potential. We continuously change the aspect ratio to see how the interaction parameters change. The left side of Fig.\ref{Figure3} shows that when $\lambda=0.59$, on-site interaction goes from repulsive to attractive but the other interaction parameters are positive. The inter-site, partial exchange and exchange interactions are smaller than $U$ by two-three orders as shown in the right side of Fig.\ref{Figure3}. However because $U$ switches its sign, other interaction terms also become important in the dynamics of JO and MQST.  The first plot in the left of Fig.\ref{Figure5}, we see that with dipolar BEC, by changing the aspect ratio of the trap one can get a transition from Rabi to Josephson regime in the small-amplitude oscillations limit, because of the on-site interaction switches its sign due to the confinement of the trap. To study the transition from JO to MQST, we choose the initial phase-difference between two condensates $\theta(0)=0$. In the second plot of Fig.\ref{Figure5} we see that when $\lambda\le0.28$ we get a transition from JO to MQST regime. The transition point depends on the value of $\phi$. 

\begin{figure}[!htbp]
\centering
\vspace{.25in}
\begin{tabular}{@{}cccc@{}}
\hspace{-0.2in}
\includegraphics[height=2.2in, width=3in]{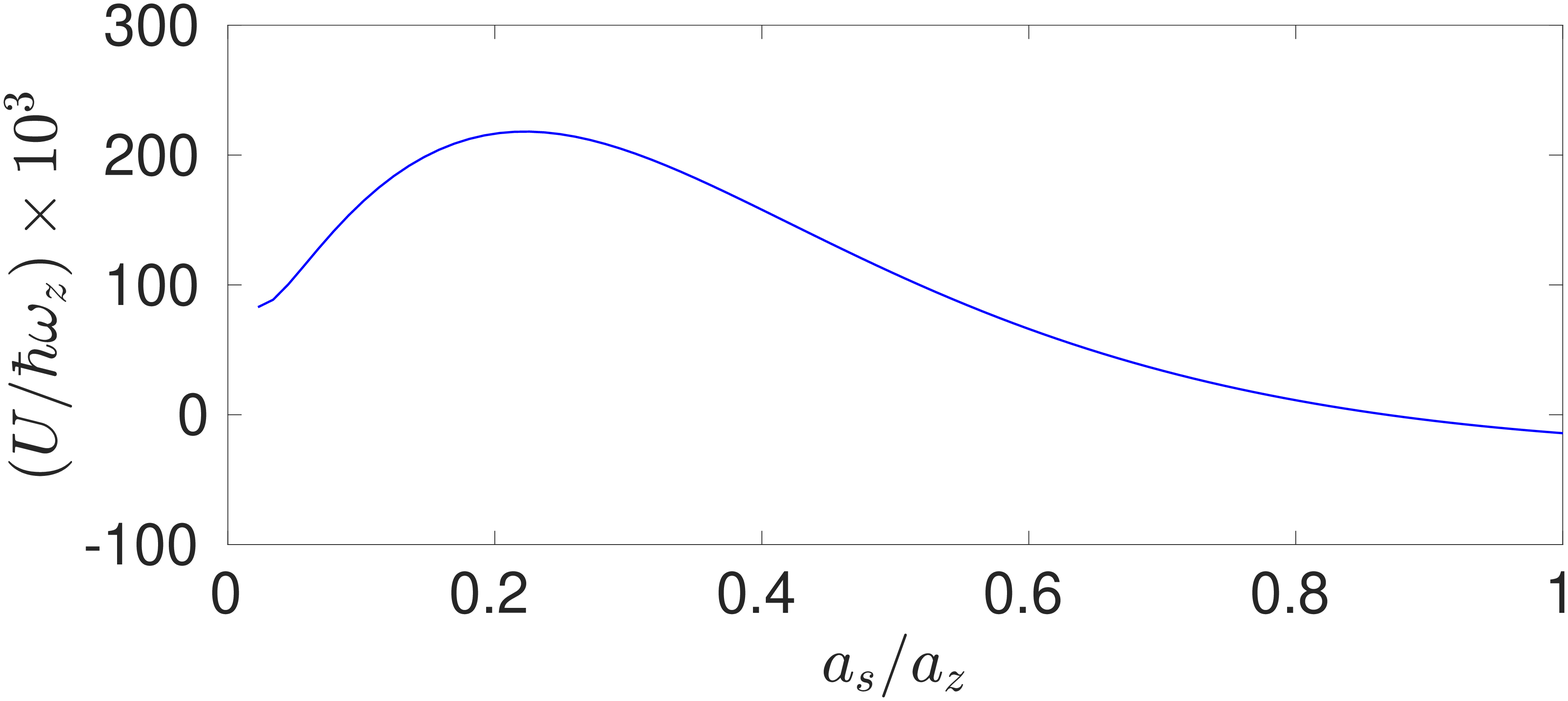} &
\includegraphics[height=2.2in, width=3in]{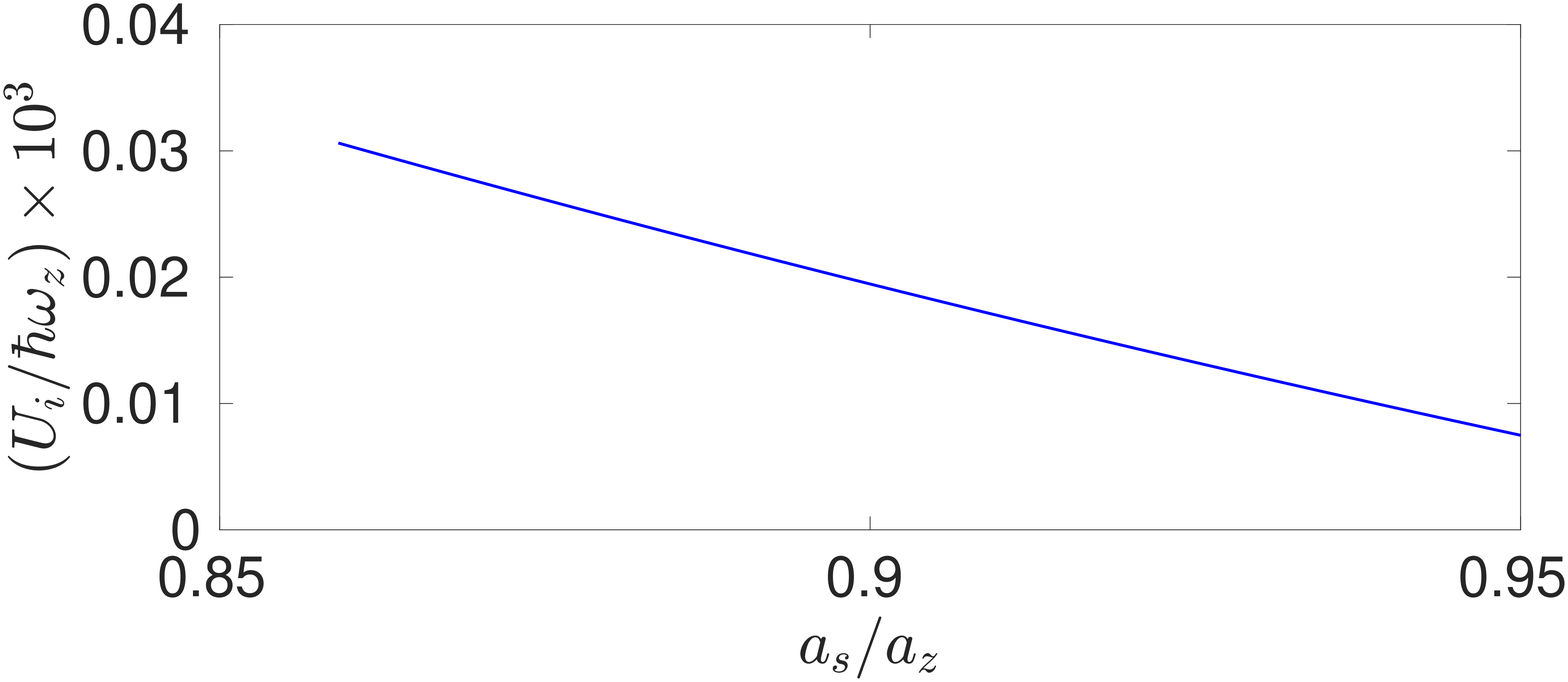}\\
\end{tabular}
\caption{\small Variation of $U$ (left) and $U_{i}$ (right) as a function of $a_{s}$ (in unit of $a_{z}$) for $\kappa a_{z}=1.2$, $r_{0}=0.006a_{z}$ and $\lambda=0.625$. $U_{i}$ is plotted where $U$ crosses zero.}
\label{Figure7}
\end{figure}
\begin{figure}[!htbp]
\centering
\vspace{.25in}
\begin{tabular}{@{}cccc@{}}
\hspace{-0.2in}
\includegraphics[height=2.2in, width=3in]{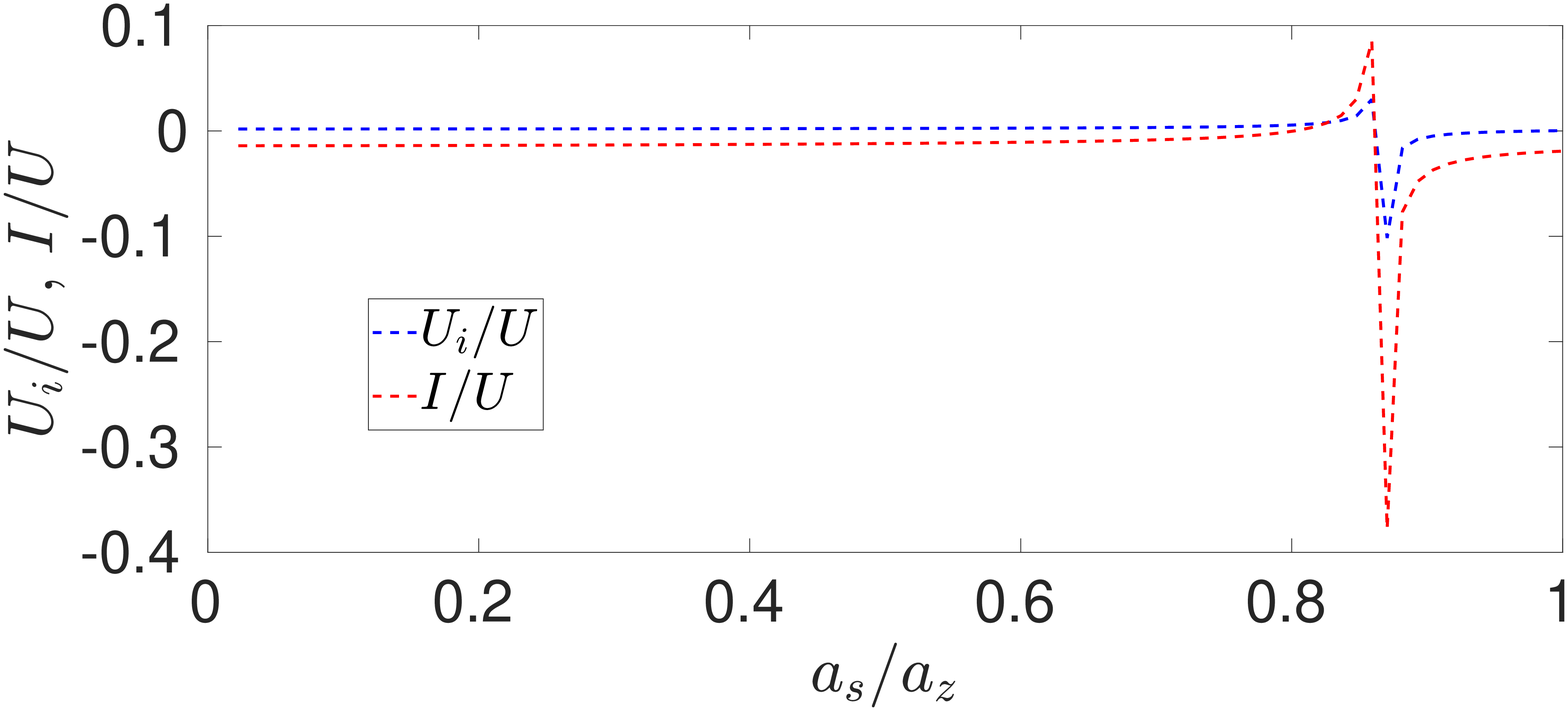} &
\includegraphics[height=2.2in, width=3in]{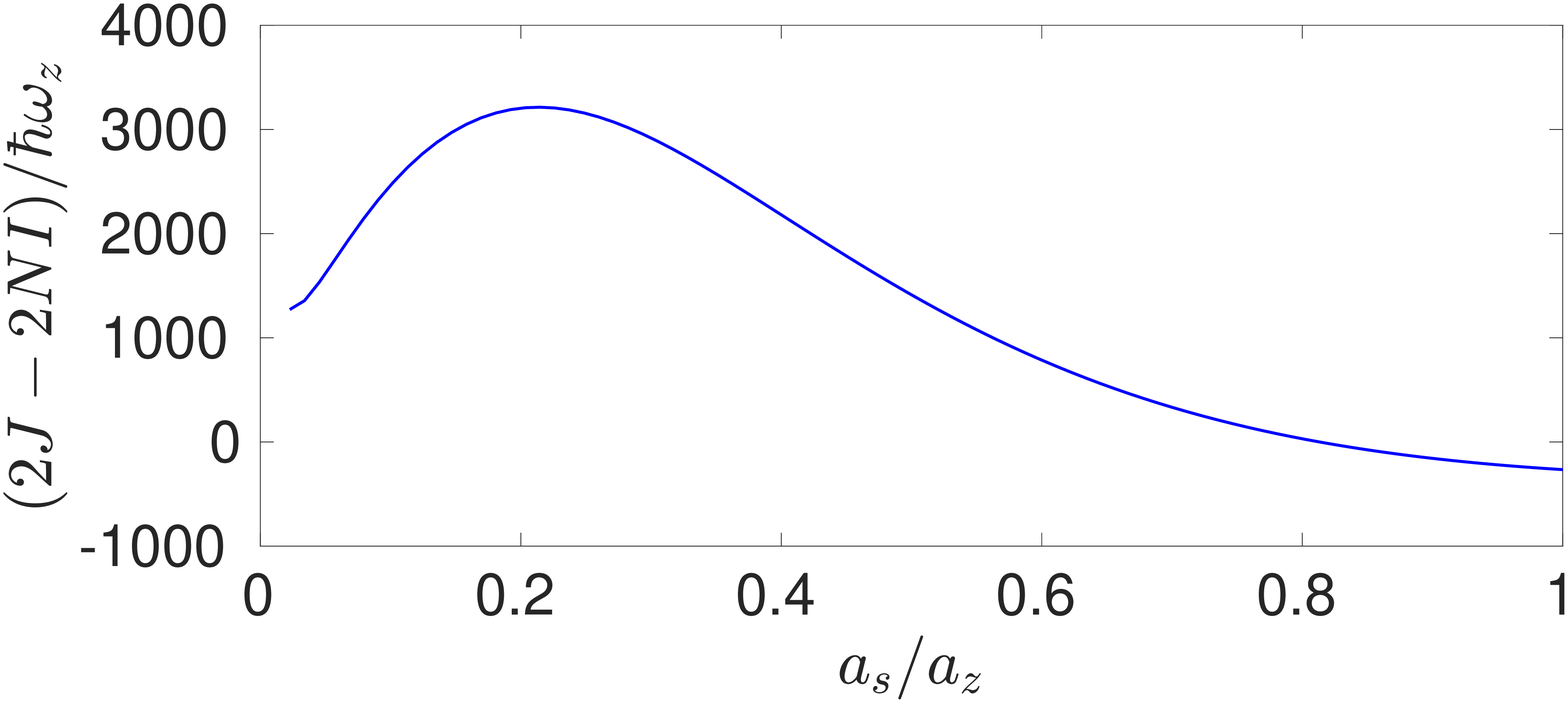}\\
\end{tabular}
\caption{\small Variation of $\frac{U_{i}}{U}$ and $\frac{I}{U}$ (left) and $(2J-2NI)$ (right) as a function of $a_{s}$  (in unit of $a_{z}$) for $\kappa a_{z}=1.2$, $r_{0}=0.006a_{z}$ and $\lambda=0.625$.}
\label{Figure8}
\end{figure}
\subsection{Finite-range interaction}\label{5:jost}
Next, to study the effects of relatively large scattering length and effective range of interaction we consider the finite-range interaction potential of Jost and Kohn \cite{Joseph-finite-12} as given in Appendix A. The dependence of $U$ on $a_{s}$ and other parameters of the Jost-Kohn potential is discussed in some detail in the Appendix A, where we calculate $U$ using the wave-functions of interacting two-particle system in a harmonic well and compare it with that calculated using the single-particle wave-functions of non-interacting system. As shown in the Appendix, the results for non-interacting case is underestimated by about one third of the results for interacting case. For numerical illustration of the dynamics of BJJ at relatively large scattering length, we fix the DW axis frequency $\omega_z=2\pi\times 85$ Hz and the radial frequency $\omega_{\rho}=2\pi\times 136$. So, $\lambda=0.625$ then we fix the value of effective range $r_{0}=0.006a_{z}$ and $\kappa a_{z}=1.2$. The left side of Fig.\ref{Figure7} shows that when scattering length nearly equals to the axial size of trap, on-site interaction energy changes from positive to negative value where the effective range is smaller by three orders of magnitude than axial size of the trap. Here we calculate the on-site interaction $U$ using the wave-functions of non-interacting two particle system in the DW. The inter-site interaction remains non-zero where the on-site interaction crosses zero as shown in the right side of Fig.\ref{Figure7}. In order to maintain the two-mode approximation, we choose the small value of the repulsive and attractive on-site interaction energy near zero value. In Fig.\ref{Figure8} we show the variation of $U_{i}/U$, $I/U$, $2J-2NI$ as a function of positive scattering length.

\begin{figure}[h]
\centering
\vspace{.25in}
\hspace{-0.3in}
\includegraphics[height=2.2in, width=3in]{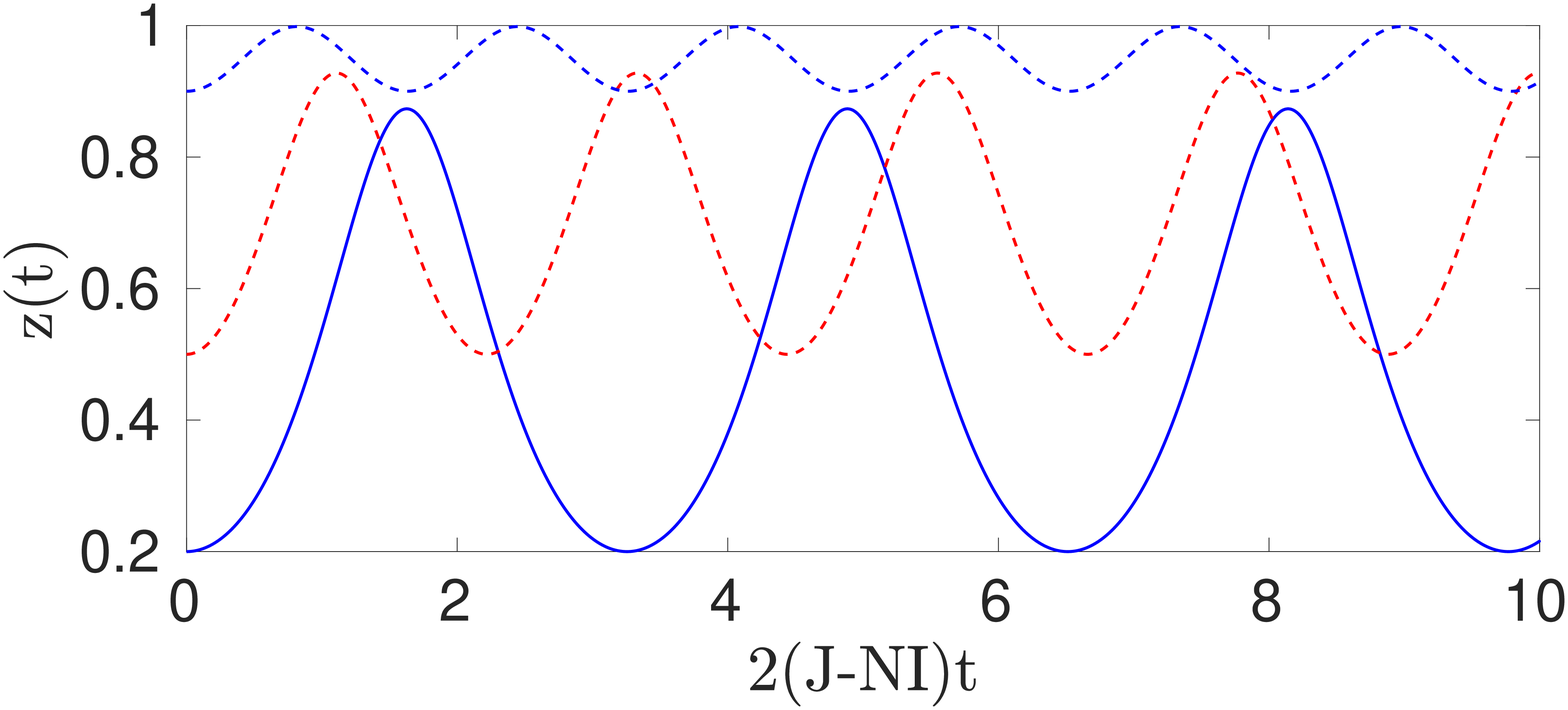} 
\caption{\small Variation of the population imbalance $z(t)$ as a function of dimensionless time $2(J-NI)t$ for different initial population imbalances z(0)=0.2 (solid blue), z(0)=0.5 (dashed-red), z(0)=0.9 (dashed-blue) in zero-phase mode for $M=-4.06$.}
\label{Figure9}
\end{figure}

\begin{figure}[h]
\centering
\vspace{.25in}
\begin{tabular}{@{}cccc@{}}
\hspace{-0.3in}
\includegraphics[height=2in, width=2.5in]{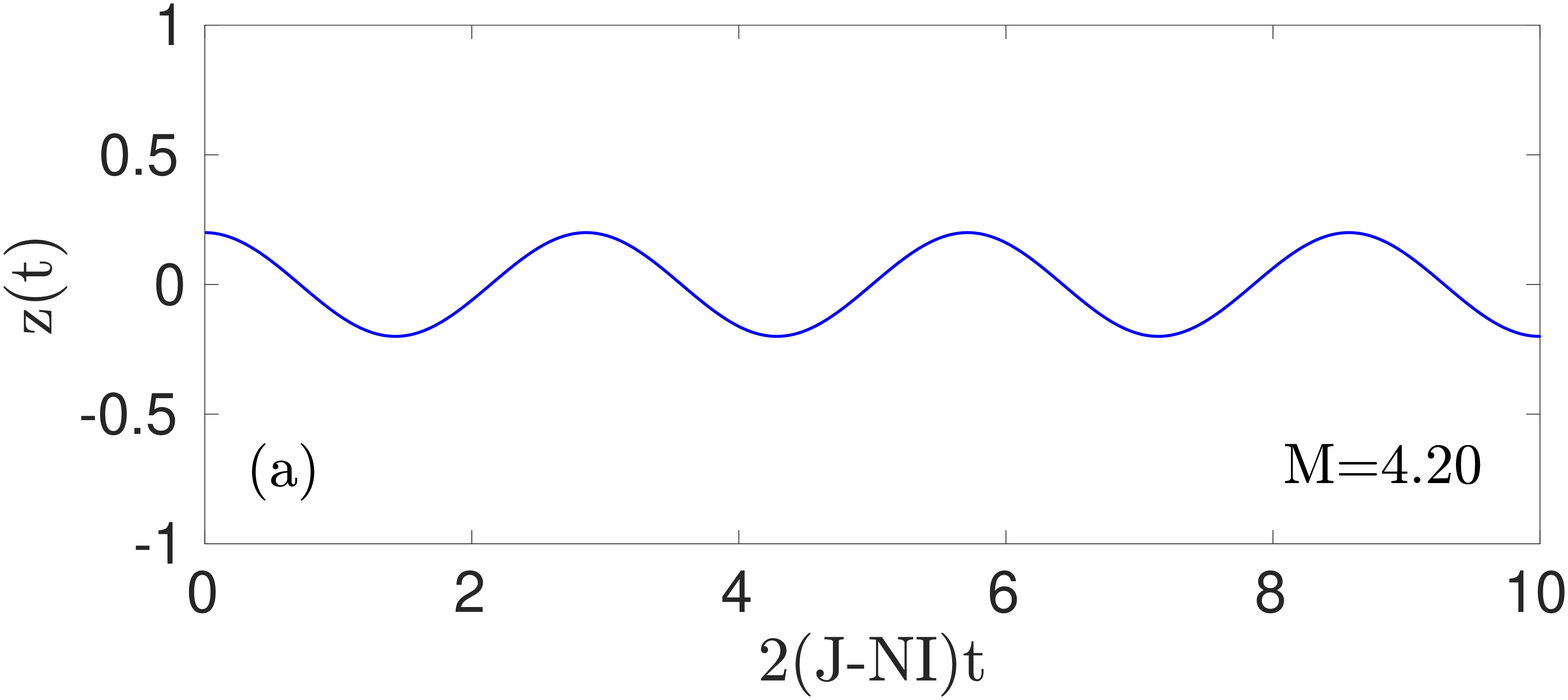} &
\hspace{-0.3in}
\includegraphics[height=2in, width=2.5in]{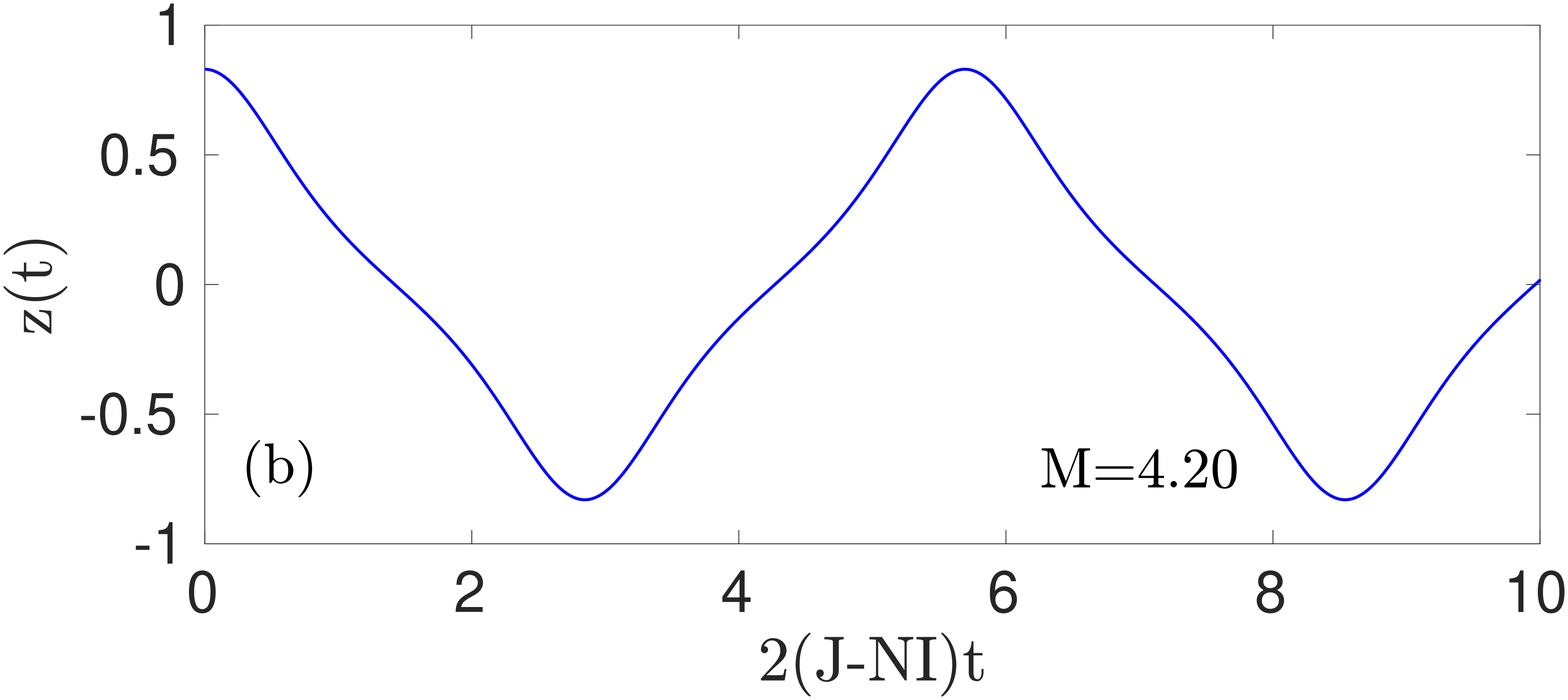} &
\hspace{-0.3in}
\includegraphics[height=2in, width=2.5in]{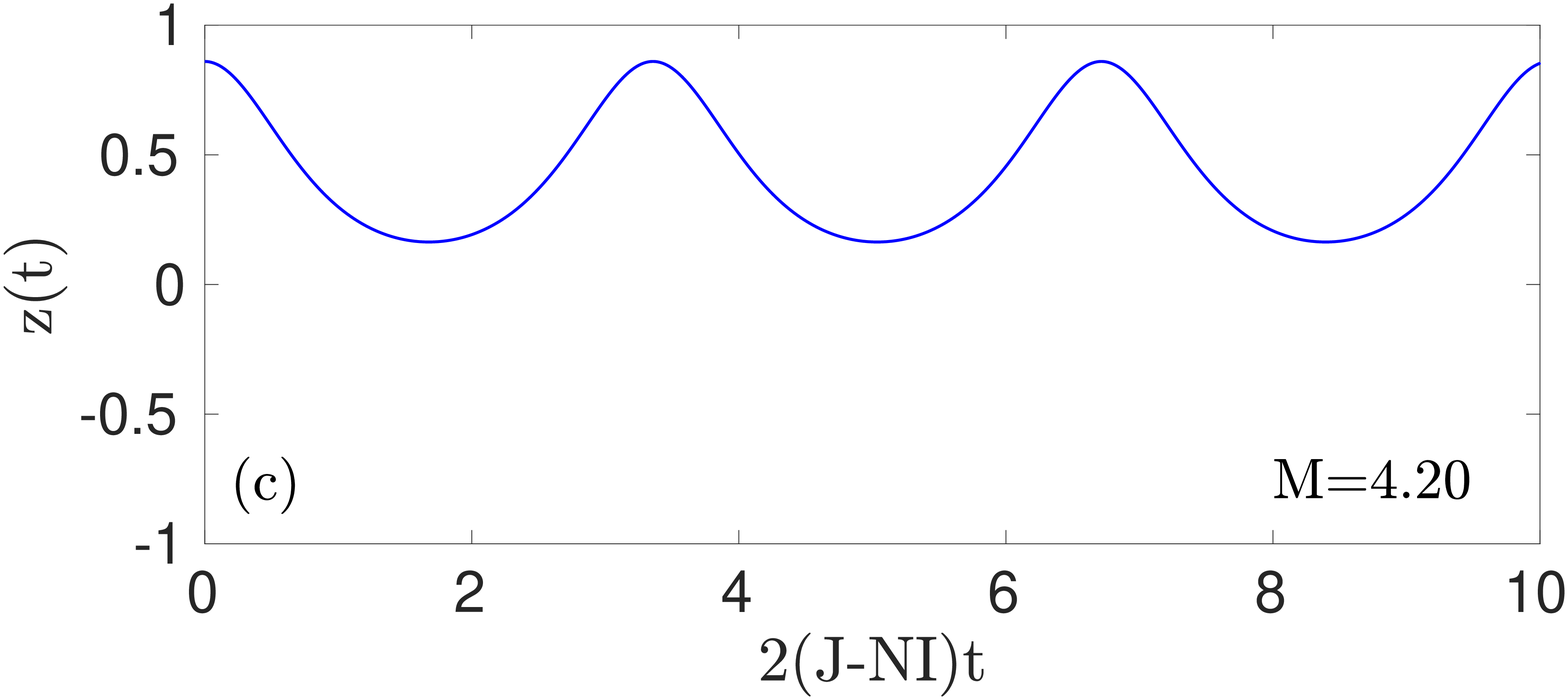}\\
\end{tabular}
\caption{\small Variation of $z(t)$ as a function of dimensionless time $2(J-NI)t$ for different initial population imbalances (a) z(0)=0.2, (b) z(0)=0.83 and (c) z(0)=0.86 in zero-phase mode for attractive on-site interaction.}
\label{Figure10}
\end{figure}

\begin{figure}[ht]
\centering
\vspace{.25in}
\begin{tabular}{@{}cccc@{}}
\hspace{-0.3in}
\includegraphics[height=2in, width=2.5in]{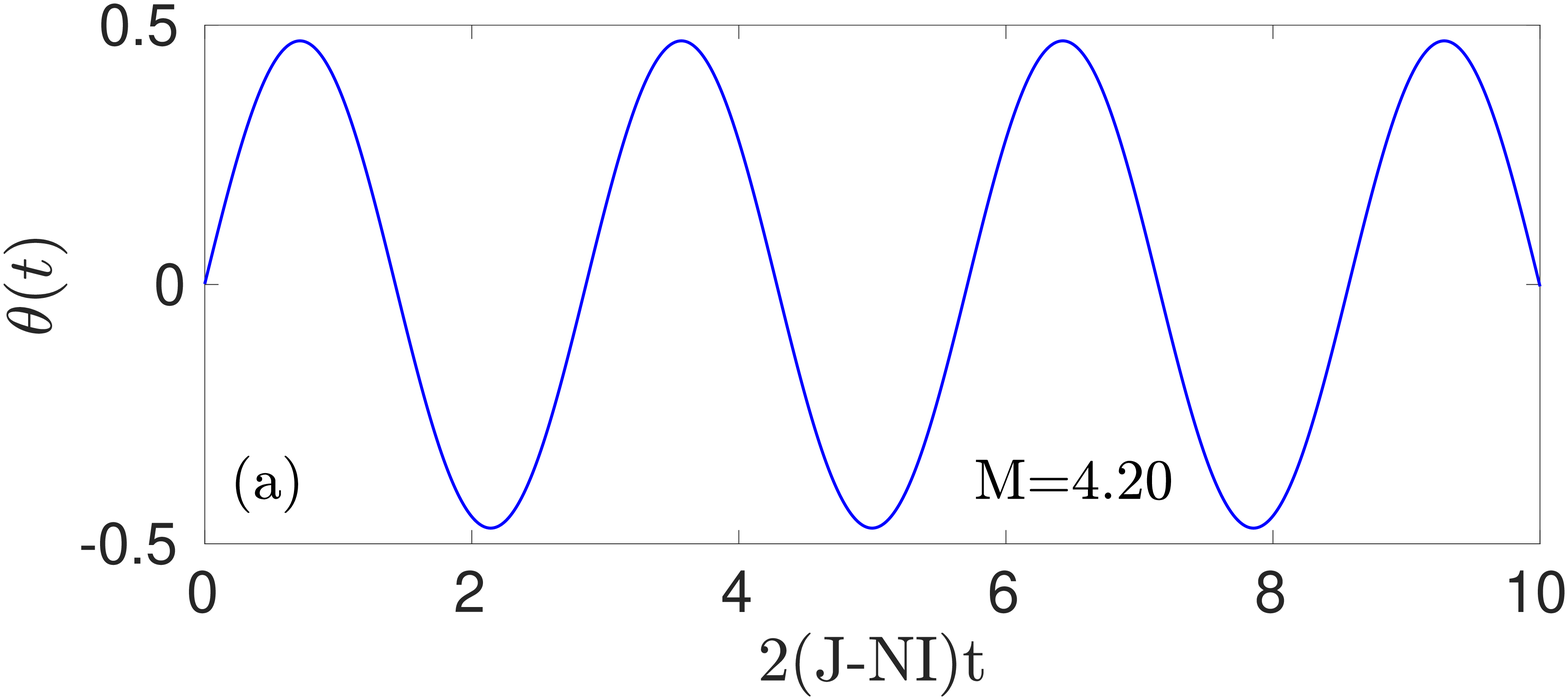} &
\hspace{-0.3in}
\includegraphics[height=2in, width=2.5in]{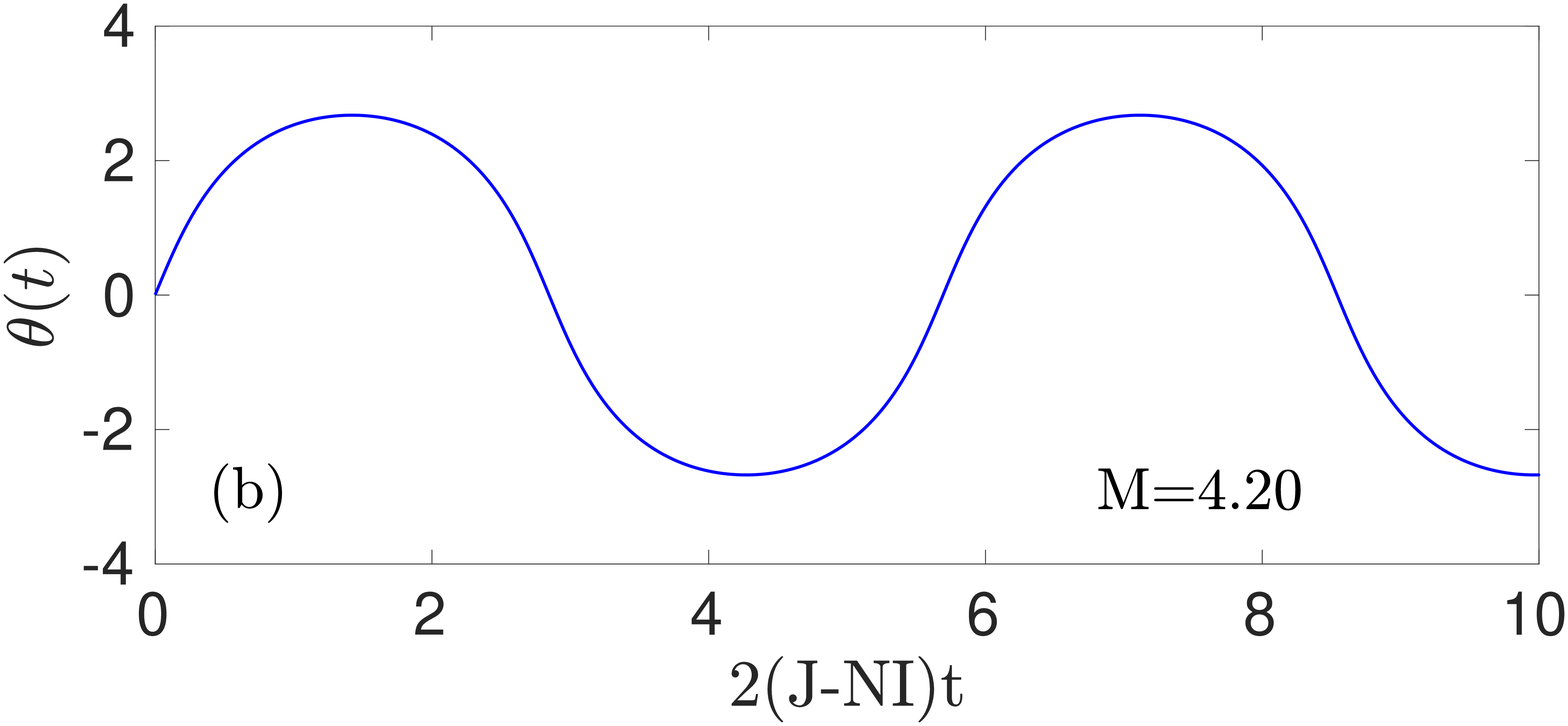} &
\hspace{-0.3in}
\includegraphics[height=2in, width=2.5in]{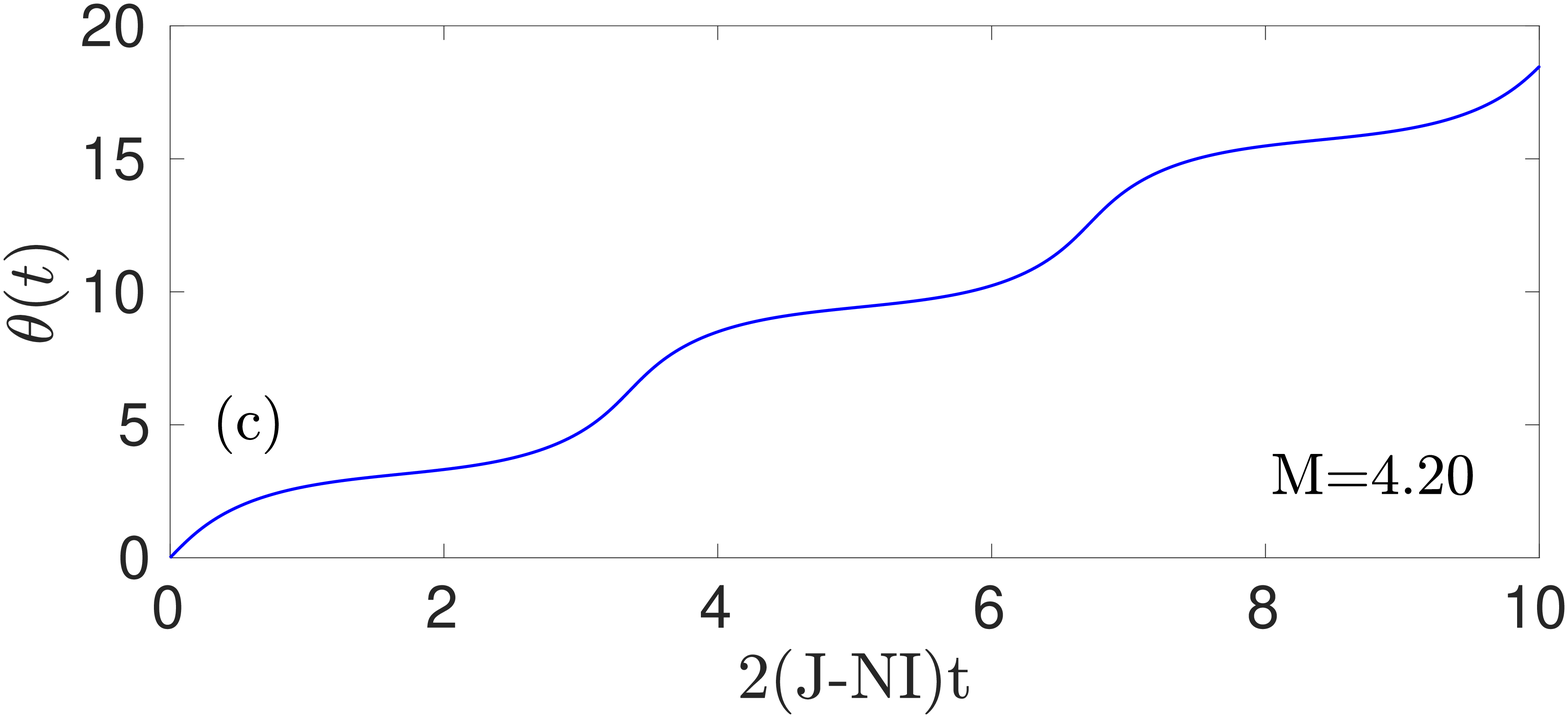}\\
\end{tabular}
\caption{\small Variation of the phase difference $\theta(t)$ as a function of dimensionless time $2(J-NI)t$ for different initial population imbalances (a) z(0)=0.2, (b) z(0)=0.83 and (c) z(0)=0.86 in zero-phase mode for attractive on-site interaction.}
\label{Figure11}
\end{figure}

The zero-phase mode oscillations describe the inter-well atomic tunneling dynamics with vanishing time-averaged value of the phase across the junction, $\theta(t)=0$. For repulsive on-site interaction, we calculate the parameters $NU=0.73$ $\hbar\omega_z$, $NU_{i}=0.03$ $\hbar\omega_z$, $NI=0.11$ $\hbar\omega_z$, $NK=-1.8\times10^{-4}$ $\hbar\omega_z$ with $N=1000$. So the value of $M=-4.06$ and $\widetilde{M}=0.001$. For attractive on-site interaction, we have $NU=-0.86$ $\hbar\omega_z$, $NU_{i}=0.04$ $\hbar\omega_z$, $NI=0.13$ $\hbar\omega_z$,  $NK=-0.004$ $\hbar\omega_z$ with $N=1000$. Here the value of $M=4.20$ and $\widetilde{M}=0.02$. For both cases the value of tunneling $J$ is $0.024$ $\hbar\omega_z$. The stationary point $(z_{s},\theta_{s})=(0,0)$ is always a saddle point for repulsive on-site interaction but for attractive on-site interaction the point $(z_{s},\theta_{s})=(0,0)$ is a maximum. So, the oscillations around a stationary point are possible only for the negative on-site interaction. From Fig.\ref{Figure9} it is clear that the system remains in self-trapped state for any initial value of the population imbalance in the repulsive on-site interaction due to the term $I$. Although the terms $U_{i}$, $I$ are typically one or two orders of magnitude smaller than on-site interaction as shown in Fig.\ref{Figure8}, their collective contributions due to condensate have nontrivial effects on the oscillations of the population imbalance. From Fig.\ref{Figure10}, we observe that for small population imbalance it oscillates around zero value. An increase of the initial population imbalance $z(0)$ adds higher harmonics to the sinusoidal oscillations. The oscillation period of the population imbalance $z$ increases with increasing $z(0)$ until, at a certain critical population imbalance $z(0)=0.86$, the oscillation is suppressed and the system is self-trapped with the phase difference between two BEC's in the left and right well evolves unbound as shown in Fig.\ref{Figure11}.    
 

In addition to zero-phase mode and MQST as discussed above, BJJ has another rich class of tunneling dynamics in which the system evolves with time-averaged phase value of $\theta=\pi$. For repulsive on-site interaction, $(z_{s},\theta_{s})=(0,\pi)$ is a minimum point whereas for attractive on-site interaction it is a saddle point. So, in the $\pi$ phase mode, the system remains self-trapped for any value of the on-site attractive interaction owing to the symmetry breaking of population imbalance. As a result, we get two types of MQST characterized by the time-averaged value of population imbalance $z <|z_{s}|\neq 0$ and $z >|z_{s}|\neq 0$ with $z_{s}$ being the stationary value of $z$ at which the symmetry breaking occurs.

\begin{figure}[!htbp]
\centering
\vspace{.25in}
\begin{tabular}{@{}cccc@{}}
\hspace{-0.2in}
\includegraphics[height=2.2in, width=3in]{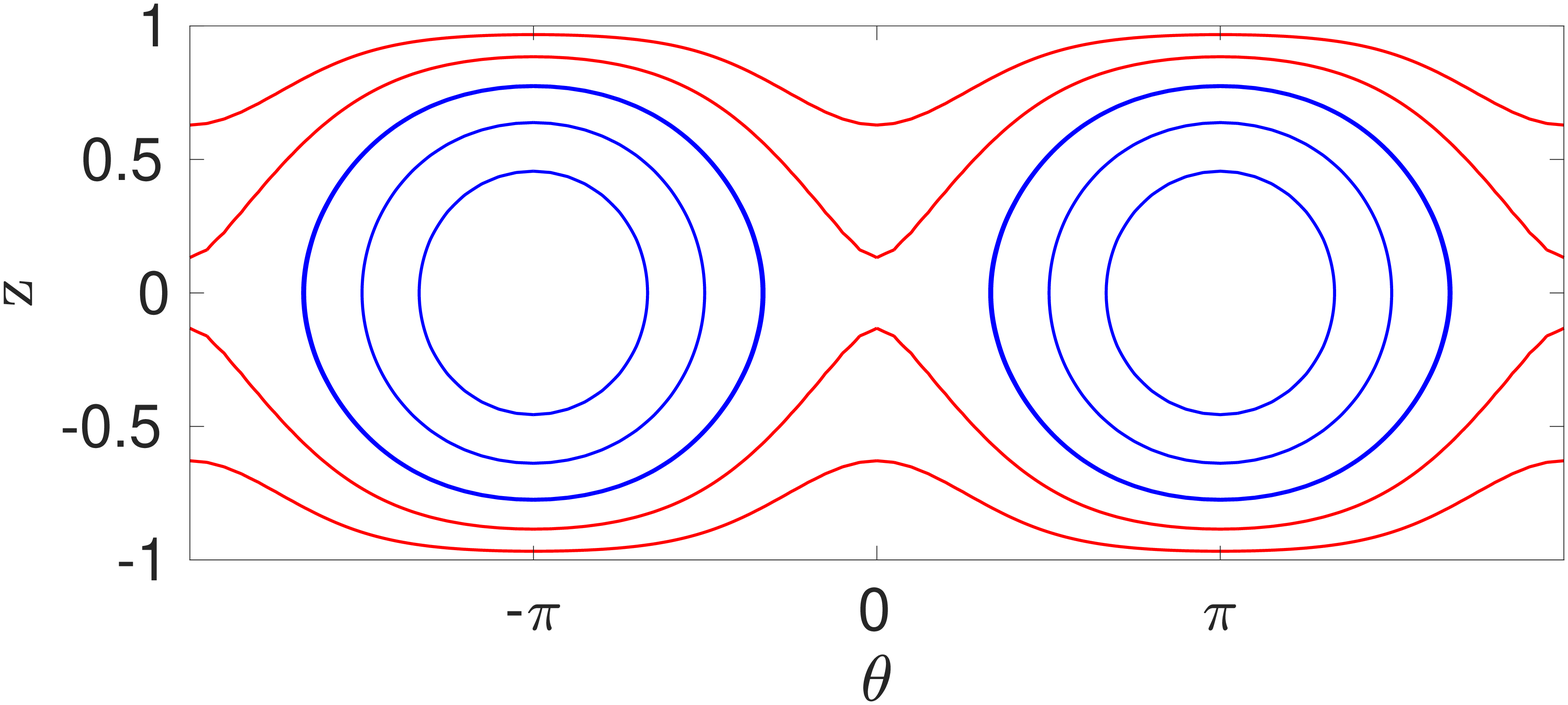} &
\includegraphics[height=2.2in, width=3in]{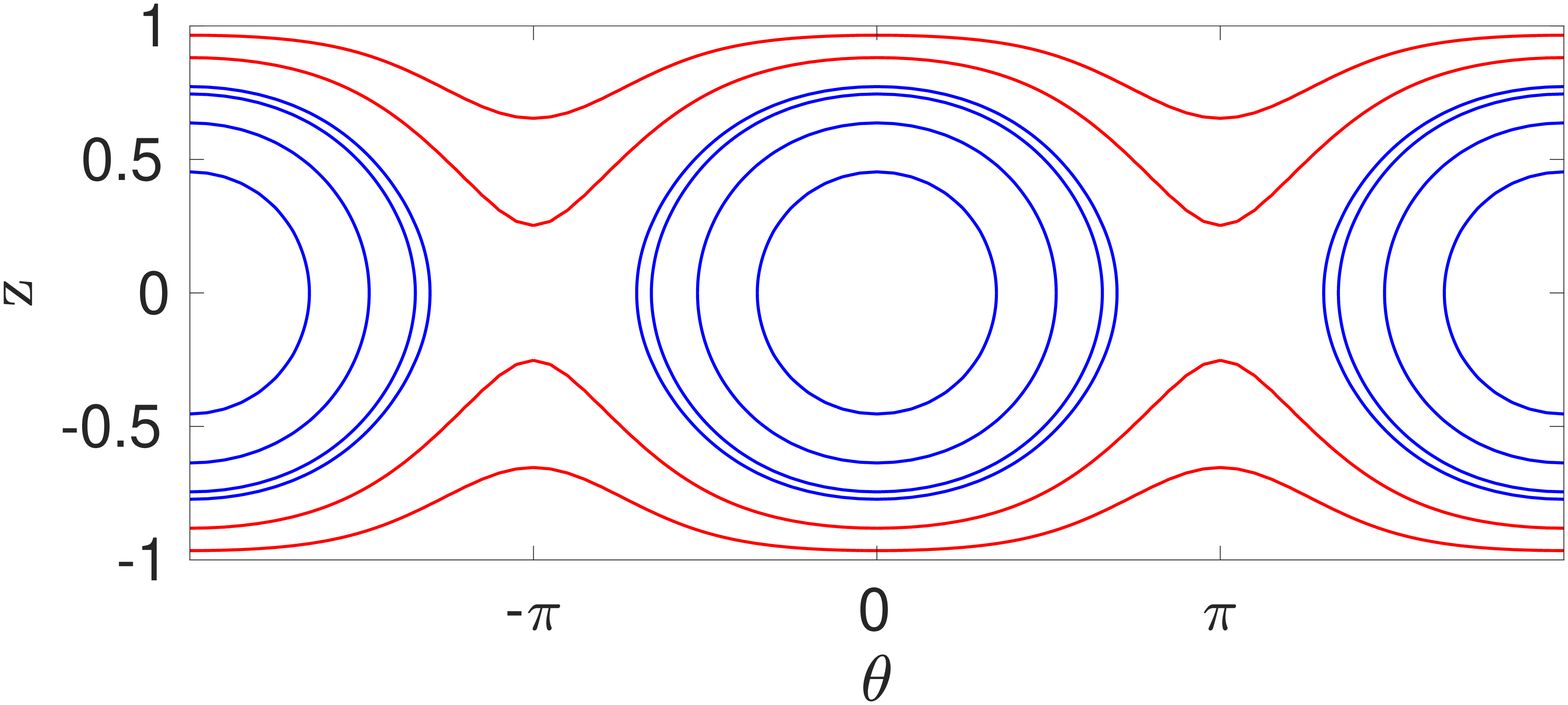}\\
\end{tabular}
\caption{\small Phase-plane portrait of the BJJ for finite-range interaction for $M=-4.06$ (left) and $M=4.20$ (right).}
\label{Figure12}
\end{figure}

The various regimes of the finite-range BJJ discussed above can be summarized in terms of phase-plane portrait, where constant energy lines are plotted in $z$-$\theta$ diagram. Fig.\ref{Figure12} shows the phase-plane plot for $M=-4.06$ and $M=4.20$, obtained by numerically solving the coupled differential Eq.(\ref{e1}) and (\ref{e2}). The first plot in the left of Fig.\ref{Figure12} describes that all trajectories with initial value of the phase difference $\theta(0)=0$ are self trapped (red lines) even for small population imbalances. The situation changes for $\theta(0)=\pi$, where for small population imbalances the trajectories are closed with no self-trapping (blue lines) but for higher population imbalance $z(0)=0.88$, the system is self-trapped. In the second plot, the phase-plane diagram is shown for $M=4.20$. In this plot when the initial value of the phase difference $\theta(0)=0$, the system oscillates around zero value for small population imbalances, given by the closed energy lines. When the initial population imbalance increases above threshold $z(0)=0.86$, the system goes to self-trapped state and for higher value of $z(0)$ the system remains always self-trapped. The system also remains always self-trapped when the initial value of phase difference $\theta(0)=\pi$.

\section{Conclusions}\label{4}
In conclusion, we have shown that, it is possible to induce transitions from JO to MQST by changing the aspect ratio of the trap keeping the scattering length fixed. For DDI, we have shown the possibility of the Rabi to Josephson transitions in small-amplitude oscillations. It is also possible to get JO and MQST in DDI by changing the aspect ratio. We have shown that the finite-range of interaction has significant influence on JO and MQST. Besides, we have shown that it is possible to get attractive interaction for positive scattering length when the scattering length is comparable to the axial size of the trap.  Josephson physics with resonant interactions is yet to be experimentally explored. The major obstacle towards such possibility will be the formation of molecules due to three-body effects and consequent loss of atoms or explosion of the condensate. But an interesting question may be posed as to what happens if the loss can be mitigated by tight confinement. Then there may arise an intriguing possibility of new Josephson physics in resonantly coupled atomic and molecular BEC's. It may be interesting to study how the coherence \cite{atom:molecule,atom:molecule1,molecule:bec} in an atom-molecular coupled BEC can affect the JO and MQST. In such atom-molecule coupled systems, the effects of trap-confinement and finite-range of interactions as studied in this paper will become important.

\section*{Acknowledgments}
One of us (SM) is thankful to the Council of Scientific and Industrial Research (CSIR), Govt. of India, for support. KA, KRD and BD are thankful to the Department of Science \& Technology, Govt. of India, for support under the the project No. SB/S2/LOP-008/2014.

\appendix
\section{Jost-Kohn potential}
The form of the Jost-Kohn potential for positive $s$-wave scattering length $a_{s}$ \cite{Joseph-finite-12} is a three parameter potential with the parameter being $a_{s}$, the effective range $r_{0}$ and another parameter $\Lambda$ which is related to the binding energy of the last bound state close to the threshold of the actual two-body interaction potential. The $s$-wave binding energy is $E_{b}=-\hbar^2\kappa^2/2\mu$ ($\kappa>0$) where $\mu$ is the reduced mass and
\begin{eqnarray}
 \kappa=\frac{1}{r_{0}}[1+\alpha]\frac{1+\Lambda}{1-\Lambda}\nonumber
\end{eqnarray}
where $-1<\Lambda<1$, $\alpha=\sqrt{(1-\frac{2r_{0}}{a_{s}})}$ and $a_{s}>2r_{0}$ for $r_{0}>0$. In terms of $a_{s}$, $r_{0}$, $\Lambda$, the potential is  
\begin{eqnarray}
V_{+}(r) &=& e^{-2(1-\alpha)\frac{r}{r_0}}\frac{8\alpha\hbar^2}{\mu r_0^2}\Bigg\{(1+\alpha\Lambda)^2(\alpha+\Lambda)^2(\alpha-1)^2(1-\Lambda^2e^{-(1+\alpha)\frac{2r}{r_0}})^2\nonumber\\ &-& \Lambda^2(1+\alpha)^2\left[(1+\Lambda\alpha)^2e^{-\frac{2\alpha r}{r_0}} - (\alpha+\Lambda)^2e^{-\frac{2r}{r_0}} \right]^2\Bigg\}\nonumber\\&\times&\Bigg\{(1+\alpha\Lambda)^2(\alpha+\Lambda^2e^{-2(\alpha+1)\frac{r}{r_0}}) - (\alpha+\Lambda)^2(e^{-2(1-\alpha)\frac{r}{r_0}} + \alpha\Lambda^2 e^{-\frac{4r}{r_0}}) \Bigg\}^{-2}
\end{eqnarray}
The idea is that, as shown in section \ref{5:jost}, even if the scattering length is positive, one can generate attractive interaction due to confinement induced effects in a finite-range interaction. This is unlike the contact interaction where the sign of the scattering length determines the nature of interaction.

\begin{figure}[!htbp]
\centering
\vspace{.25in}
\begin{tabular}{@{}cccc@{}}
\hspace{-0.2in}
\includegraphics[height=2.2in, width=3in]{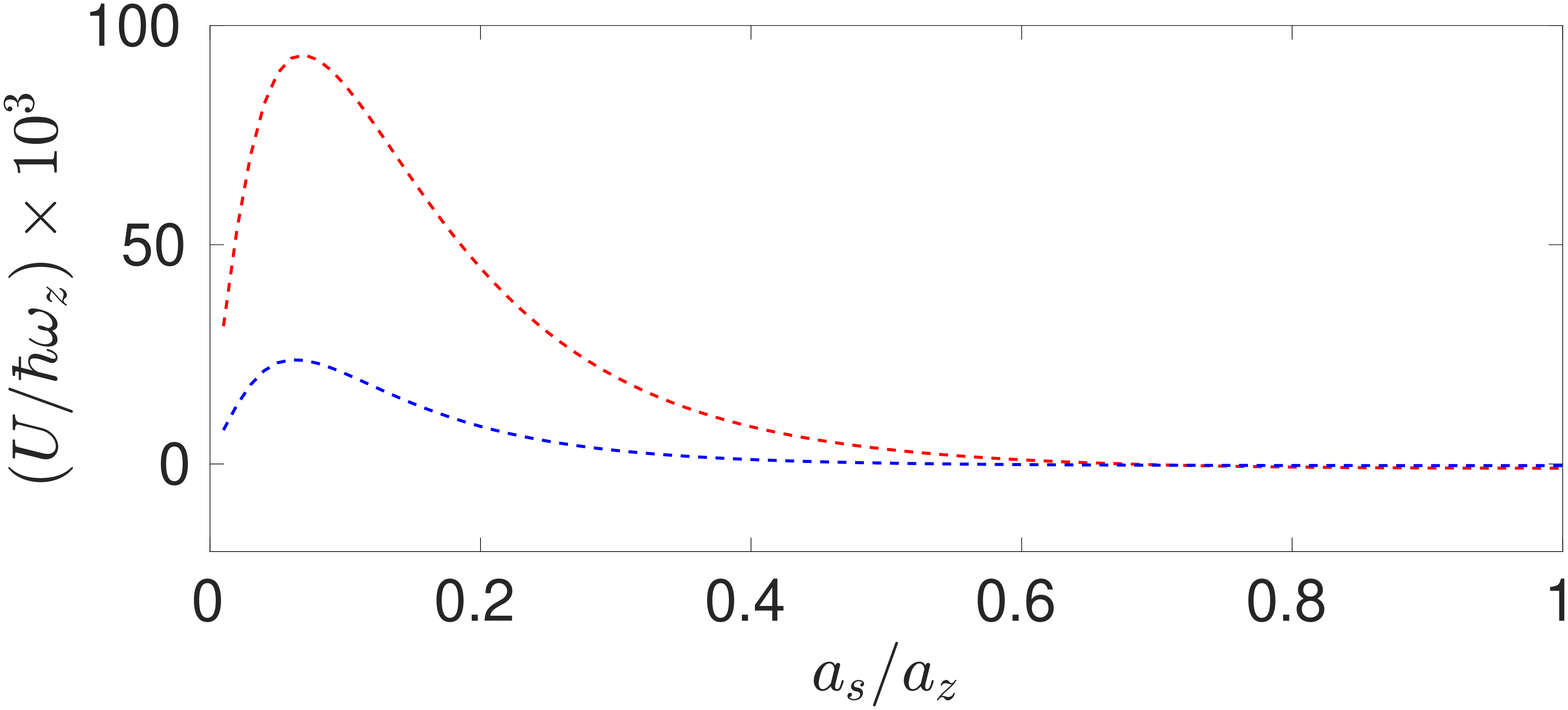} &
\includegraphics[height=2.2in, width=3in]{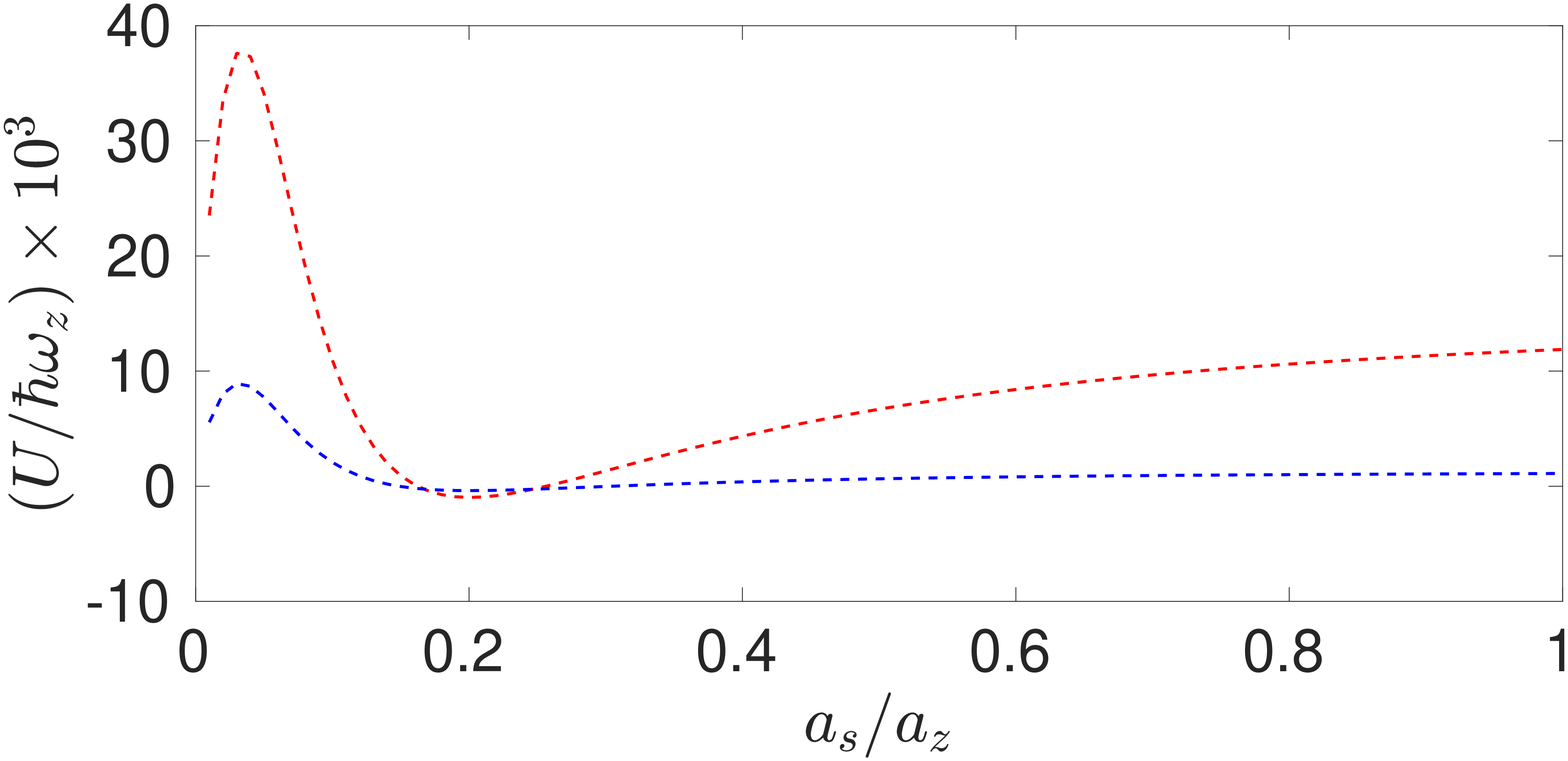}\\
\end{tabular}
\caption{\small Variation of $U$ (in unit of $\hbar\omega_z$) as a function of $a_{s}$ (in unit of $a_{z}$) for $r_{0}=0.001a_{z}$, $\lambda=0.625$ and $\kappa a_{z}=1$ (left), $\kappa a_{z}=5$ (right) for interacting (dashed-red) and non-interacting (dashed-blue) cases.}
\label{Figure13}
\end{figure}

\begin{figure}[!htbp]
\centering
\vspace{.25in}
\begin{tabular}{@{}cccc@{}}
\hspace{-0.2in}
\includegraphics[height=2.2in, width=3in]{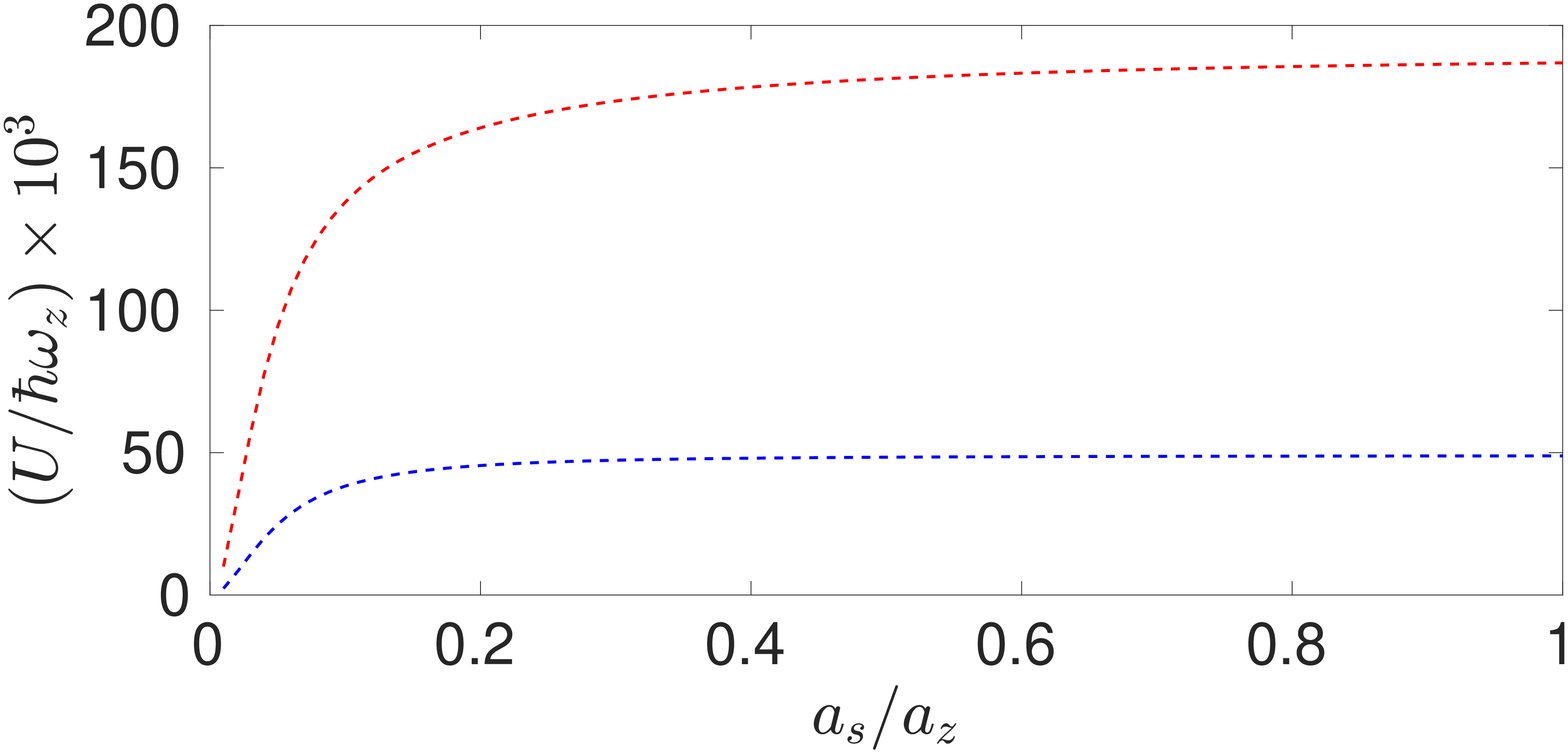}
\end{tabular}
\caption{\small Variation of $U$ (in unit of $\hbar\omega_z$) as a function of  $a_{s}$ (in unit of $a_{z}$) for $\kappa a_{z}=500$, $r_{0}=0.001a_{z}$, $\lambda=0.625$ for interacting (dashed-red) and non-interacting (dashed-blue) cases.}
\label{Figure15}
\end{figure}

\begin{figure}[!htbp]
\centering
\vspace{.25in}
\begin{tabular}{@{}cccc@{}}
\hspace{-0.2in}
\includegraphics[height=2.2in, width=3in]{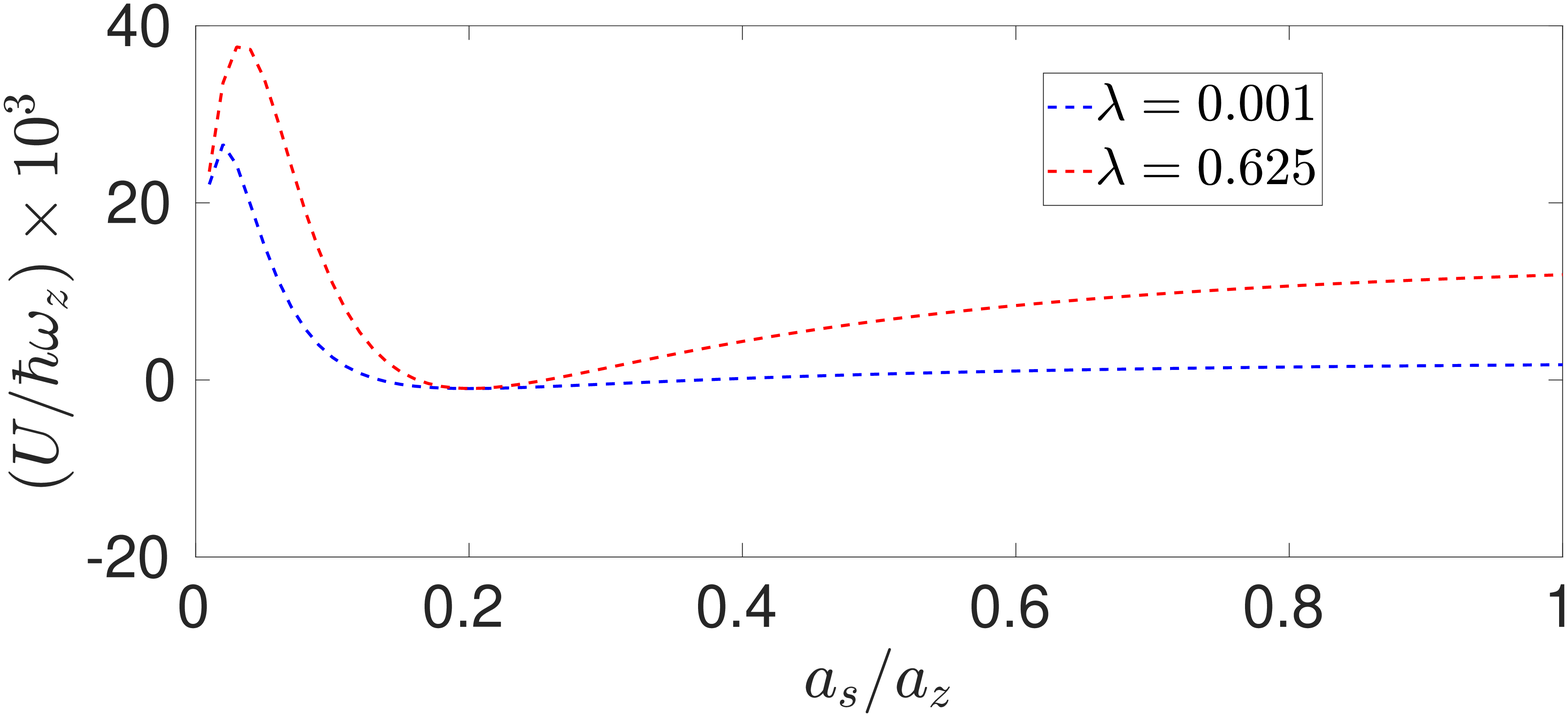} &
\includegraphics[height=2.2in, width=3in]{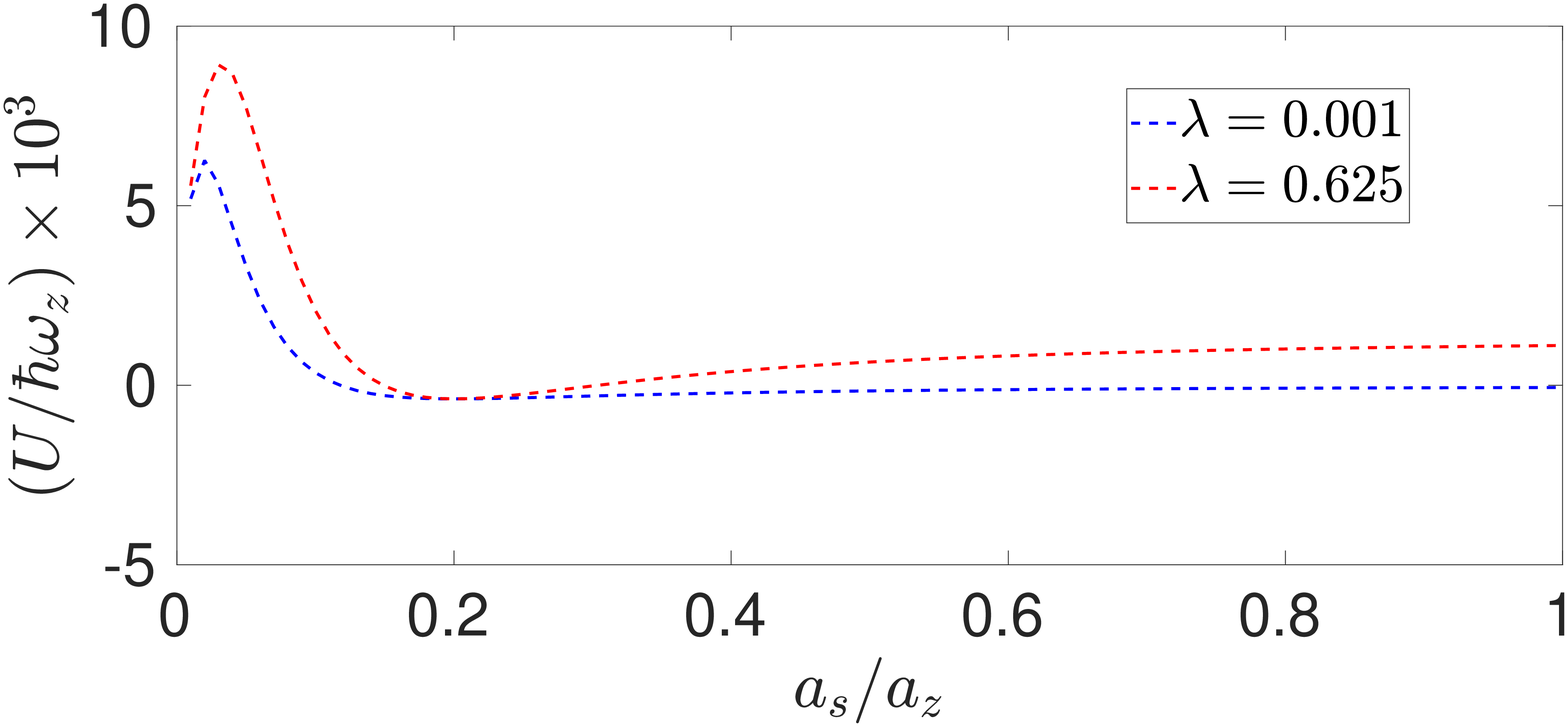}\\
\end{tabular}
\caption{\small Variation of $U$ (in unit of $\hbar\omega_z$) as a function of $a_{s}$ (in unit of $a_{z}$) for $\kappa a_{z}=5$, $r_{0}=0.001a_{z}$ for interacting (left) and non-interacting (right) cases for two different values of aspect ratio.}
\label{Figure16}
\end{figure}

Though our finite-range interaction potential does not support any bound state, it explicitly depends on the parameter $\kappa$. In the limit $\kappa\rightarrow\infty$, the interaction is described by scattering length $a_{s}$ and the effective range $r_{0}$ which can be obtained from the well-known effective range expansion
\begin{eqnarray}
k\cot\eta(k)=-\frac{1}{a_s}+\frac{1}{2}r_{0}k^2+... \nonumber
\end{eqnarray}
where $\eta(k)$ is the scattering phase shift associate with the wave-vector $k$. However, for small $\kappa$, the effective range expansion is modified with modified scattering length and modified effective range \cite{smal:arxiv}
\begin{eqnarray}
\bar a_s &=& a_s-\frac{2}{\kappa} \\
\bar r_{0} &=& \frac{a_s}{\bar a_s} \Bigg ( \frac{1}{\kappa} \Bigg) \Bigg[ \kappa r_0-4 +\frac{1}{2\kappa a_s}+\frac{1- 2 r_0\kappa}{4- 4 \kappa a_s}\Bigg]
\end{eqnarray}
It is clear that $\bar r_{0}$ is negative but $\bar a_{s}>0$ for $\kappa r_{0}<<1$ and $\kappa a_{s}>2$ with $r_{0}>0$. Negative effective range occurs for a narrow resonance \cite{ohara:prl:2012,hutson:2014} and may be interpreted as the breakdown of the standard effective range expansion \cite{smal:arxiv}. On the other hand $\bar a_{s}<0$ for $\kappa a_{s}<2$. Here we choose $\lambda=0.625$ then we fix the value of the effective range $r_{0}=0.001a_{z}$. We vary $\kappa$ to different values to see how the interaction parameters behave as a function of $a_{s}$. From the left side of Fig.\ref{Figure13} describes the variation of on-site interaction $U$ as a function of  $a_{s}$ when $\kappa a_{z}=1$, showing that when scattering length is large ($a_{s}=0.65a_{z}$) then on-site interaction changes from positive to negative value. Here we calculate the on-site interaction using the wave-functions of interacting (dashed-red) two-particle system in a harmonic well and compare the results with that using the wave-functions of non-interacting (dashed-blue) two particle system. Though both results are qualitatively similar, the results with the wave-functions of non-interacting system is underestimated by about one-third of that with the wave-function of interacting system as shown in Fig.\ref{Figure13} for which $\bar r_{0}$ and $\bar a_{s}$ are always negative. Next if we consider the value $\kappa a_{z}=5$ then from the right side of Fig.\ref{Figure13} we have a situation where $\bar a_{s}>0$ and $\bar a_{s}<0$. Here also $\bar r_{0}$ is negative always.
For larger values of $\kappa a_{z}=500$, as shown in Fig.\ref{Figure15}, the variation of $U$ as a function of $a_{s}$ is almost linear in the small scattering length limit. But in the large scattering length limit $U$ becomes constant. The values of $\bar r_{0}$ and $\bar a_{s}$ are always positive in this case. For higher values of $\kappa$, total many-body on-site interaction $NU$ becomes large compare to the energy gap between ground and first excited state of the single well leading to the breakdown of two-mode approximation. Next in Fig.\ref{Figure16} we plot the on-site interaction energy as a function of $a_{s}$ for two different values of aspect ratio $\lambda$. When $\lambda=0.001$, we have quasi-1D regime for which $U$ becomes saturated in the large scattering length limit when $\kappa a_{z}=5$. When $\lambda=0.625$, the value of $U$ is larger compare to the quasi-1D regime. So, to study the BJJ at large scattering length we have to choose the values of $\kappa a_{z}$ small so that our $NU$ lies well below the gap between ground and first excited state of the single well.        

\end{document}